\pdfoutput=1
\documentclass[preprint,number,sort&compress,5p,twocolumn,times]{elsarticle}

\usepackage{amsmath}
\usepackage{multirow}
\usepackage{subfig}
\usepackage{listings}
\usepackage{float}
\usepackage[linesnumbered,lined,commentsnumbered]{algorithm2e}
\usepackage[hidelinks]{hyperref}

\newfloat{lstfloat}{htbp}{lop}
\floatname{lstfloat}{Listing}

\lstdefinestyle{mystyle}{
    backgroundcolor=\color{white},
    commentstyle=\color{green!50!black},
    keywordstyle=\color{blue},
    numberstyle=\tiny\bf,
    stringstyle=\color{red!50!blue},
    basicstyle=\ttfamily\scriptsize,
    breakatwhitespace=false,         
    breaklines=true,                 
    captionpos=b,                    
    keepspaces=true,                 
    numbers=left,                    
    numbersep=5pt,                  
    showspaces=false,                
    showstringspaces=false,
    showtabs=false,                  
    tabsize=2,
    columns=flexible
}
 
\newcommand{\ra}{\ensuremath{\rightarrow}}

\SetKwProg{Fn}{Function}{}{}
\SetKw{Continue}{continue}

\makeatletter
\newcommand{\removelatexerror}{\let\@latex@error\@gobble}
\makeatother

\usepackage{tikz}
\usepackage{standalone}
\usepackage{marvosym}

\usetikzlibrary{decorations.pathreplacing}
\usetikzlibrary{patterns}
\usepackage{footmisc}

\usepackage{amssymb}

\usepackage[shadow,backgroundcolor=green!40]{todonotes}

\journal{Information Systems}
\usepackage{color}

\begin{document}

\begin{frontmatter}

\title{Event-based Failure Prediction in Distributed Business Processes}

\author[tuv]{Michael Borkowski}
\author[uvi]{Walid Fdhila}
\author[uro]{Matteo Nardelli}
\author[uvi]{Stefanie Rinderle-Ma}
\author[tuv]{Stefan Schulte}

\address[tuv]{Distributed Systems Group, TU Wien}
\address[uro]{University of Rome Tor Vergata}
\address[uvi]{Workflow Systems and Technology, University of Vienna}

\begin{abstract}
Traditionally, research in Business Process Management has put a strong focus on centralized and intra-organizational processes. However, today's business processes are increasingly distributed, deviating from a centralized layout, and therefore calling for novel methodologies of detecting and responding to unforeseen events, such as errors occurring during process runtime. In this article, we demonstrate how to employ event-based failure prediction in business processes. This approach allows to make use of the best of both traditional Business Process Management Systems and event-based systems. Our approach employs machine learning techniques and considers various types of events. We evaluate our solution using two business process data sets, including one from a real-world event log, and show that we are able to detect errors and predict failures with high accuracy.
\end{abstract}

\begin{keyword}
failure prediction \sep event-based systems \sep business process management \sep machine learning

\vspace{2mm}

\fbox{\begin{minipage}{\textwidth}
NOTICE: This is the authors' version of the manuscript accepted for publication in Information Systems. Please cite as: \textbf{Michael Borkowski, Walid Fdhila, Matteo Nardelli, Stefanie Rinderle-Ma, Stefan Schulte: Event-based Failure Prediction in Distributed Business Processes. Information Systems (2018), \url{https://doi.org/10.1016/j.is.2017.12.005}.} The content of the version layouted by Elsevier found using this DOI is identical to this paper.

\copyright\,2017. This manuscript version is made available under the CC-BY-NC-ND 4.0 license. \\
\url{http://creativecommons.org/licenses/by-nc-nd/4.0/}
\end{minipage}}

\end{keyword}

\end{frontmatter}

\section{Introduction}
\label{sec:intro}
Business Process Management (BPM) addresses the problem of how to design, analyze, configure, enact, and evaluate business processes \cite{weske12}. In the last two decades, research efforts in the BPM field have resulted in a rich toolset covering all phases of the BPM lifecycle, however, with a strong focus on centralized and intra-organizational processes. In contrast, distributed and decentralized business processes have received comparably little attention \cite{breu13}.

Nevertheless, today's business processes are to a large degree inter-organizational and distributed, since companies need to collaborate in order to generate a desired output. Examples for distributed processes can be found in healthcare, manufacturing~\cite{schulte14}, or smart grids \cite{rohjans12}, amongst others. 

One way to include a notion of distribution into business processes is by adopting basic concepts from the field of event-based systems (EBS) \cite{muehl06}. As the name implies, EBS define a software architecture pattern which is based on \emph{events}, i.e., state changes of process-related objects \cite{muehlen15}. Instead of applying a request/response, \textit{pull}-based messaging pattern, EBS decouple message producers and consumers by \textit{pushing} events to receivers. %
As one prominent example, the publish/subscribe pattern is based on events which are sent from a publisher to subscribers \cite{eugster03}. Importantly, event messages are not aimed at a particular receiver. Instead, a notification service decouples producers and consumers and delivers events whenever necessary. This allows separation of event-based communication from computation \cite{muehl06}. While EBS can also be centralized, distribution is usually seen as an inherent feature of modern EBS. This applies both to the potential distribution of data to be processed as well as the EBS itself, i.e., such a system can be distributed in order to allow horizontal or vertical scalability. With the advent of the Internet of Things (IoT) \cite{atzori10}, which is a highly distributed, worldwide network based on sensor, communication, networking, and computation technologies \cite{li15}, a virtually unlimited number of potential event sources exist.  

Events can be used to control and adapt distributed business processes during design time, change time, and runtime, or to simply exchange data between different process stakeholders. This includes events coming from IoT devices, but also data from business intelligence or any other event-producing system. 

One particular application area of events in business processes is their usage in order to predict potential failures during process execution. To the best of our knowledge, research on fault tolerance in business processes has so far focused on the exploitation of process-inherent knowledge, e.g., from process logs \cite{Kang:2012, bezerra09}, while only a limited amount of approaches consider context data like events. Nevertheless, context events, e.g., from IoT technologies or other data sources, can influence the outcome of a process, and should therefore be taken into account. Thus, the goal of this paper is to examine the exploitation of events in order to find errors and predict potential failures during (distributed) process execution. Based on this prediction, it is possible to start according countermeasures in order to prevent a fault from causing an actual failure.
The overarching research goal of this paper is approached along the following research questions: 

\begin{enumerate}
\item How can contextual data of external EBS be combined with process-intrinsic events of a (distributed) business process in order to predict process failures?
\item How to formalize and automate failure prediction in (distributed) process settings?
\item Is the approach feasible when used in a real-world scenario, and how well does the failure prediction perform?
\item What is the impact of the distribution of processes on the failure prediction?
\end{enumerate}

For answering these questions, this paper investigates the exploitation of events in business processes by combining Business Process Management Systems (BPMS) with EBS. We employ neural networks, a technique used in machine learning (ML), to predict whether a running business process is likely to fail, and at which step. Since business processes often include interactions between various partners, we analyze the impact of inter-organizational processes on the prediction performance.

The paper follows the design science methodology \cite{DBLP:books/sp/Wieringa14}, taking the following steps: First, the relevance of the problem is motivated and illustrated by an example from the logistics domain (Section~\ref{sec:background}). The main artifacts comprise the architecture of the overall solution (Section~\ref{sec:solution}) as well as the algorithm for failure prediction including performance optimization (Section~\ref{sec:machine}). The proposed algorithm is evaluated with respect to its feasibility and performance (Section~\ref{sec:evaluation}). The evaluation therefore comprises a technical part, i.e., a prototypical implementation, as well as a validation part. The latter is conducted based on two realistic data sets, i.e., one real-world data set from the finance domain, and one data set that is simulated for a real-world distributed manufacturing process. Both data sets are pre-processed in order to be usable to detect faults and predict failures of different kinds and varying amounts and levels of distribution.
In addition, we comment on the related work in Section~\ref{sec:related}. Finally, we conclude the paper in Section~\ref{sec:conclusion}.

\section{Fundamentals and Motivation}
\label{sec:background}

In this section, we provide an example for a business process involving multiple partners. This example demonstrates the motivation for predicting a failure impacting the final process outcome before the failure's occurrence, in order to allow a timely reaction to such a failure.

\textit{Example: A container with bananas is shipped from South America to Europe. This shipment is part of a supply chain business process ``Send produce from South American plantation to Viennese supermarket''. The container is equipped with sensors, which at some point of time identify a temperature exceeding limits, and therefore emit an according event. Thus, it is possible to derive that, with very high probability, the bananas are somewhat rotten and therefore cannot be shipped to the supermarket. Most importantly, this can be done \emph{before} the container is actually opened at its destination in Europe, and it is possible to ship another container (as a countermeasure to overcome the faulty process/shipment).}

In the following subsections, we present some technical fundamentals and preliminaries helpful to understand the contributions of this paper.

\subsection{Process Collaborations}
\label{sub:choreo}
As we are dealing with both intra- and inter-organizational processes, this section provides a brief overview on differences between these two concepts. 
While a process orchestration represents the business logic of a single organization, a process collaboration involves multiple organizations that collaborate to achieve a common goal \cite{BarrosDO05}. In a collaboration, we mainly distinguish between three different but overlapping layers: (i)~private processes, (ii)~public processes, and (iii)~a choreography model \cite{breu13}.

The private process represents the business logic of one organization and corresponds to its executable process, i.e., orchestration. In particular, it defines the relationship between its tasks and characterizes both its control and data flow. The internal logic as well as the corresponding data is hidden from the other organizations \cite{cs4216}. 
In contrast, a public process represents the interface with the other organizations and includes public tasks as well as the interaction activities from the perspective of one single organization. The public model logic and data can be visible to the other involved organizations.
Finally, a choreography model gives an overview of the collaboration between different organizations and defines the interaction flow between them. In particular, a choreography model describes how the latter should interact with each other, and furthermore specifies the content of the message exchanges at each step. 

In the presented example, we argue that a logistic business process includes private and public processes, as well as a choreography model: First, the initiator of the process, e.g., the owner of the Viennese supermarket, is interested in a timely and efficient shipment. Second, the shipping company, as commissioned by the shop owner, has a business process of its own (private process). In turn, the private process of the shop owner might not be visible to the shipment company in its entirety, but only the interface with the shipment company. These interfaces together constitute the public process models. Note that the difference between the public and the private model is caused by varying stakeholder interests. The total workflow involving all partners to achieve the common goal of shipping goods is the choreography model.

At runtime, each organization holds a set of process instances which run concurrently with the process instances of the collaborating organizations. Instances of different process partners are not independent, and interact with each other. In order to correlate between them, we assume a global instance identifier. The latter is generated by one organization, i.e., the initiator, and transmitted to the other organizations for each new collaboration instance. The global identifier is important for correlating between exchanged messages/data and the corresponding process instances. 

\subsection{Event Streams}

A process task is a unit of work that is performed to complete a job, and involves a set of resources, i.e., humans or machines. When a resource performs a task, data is emitted in form of events \cite{Koetter2013}. We distinguish between two main kinds of events~\cite{vonAmmon:2009,4595562,Bohmer:2016}:

\label{eventstreams}

\begin{description}
	\item[Intrinsic Events] Process steps starting and finishing generate events intrinsic to the process model. They consist of the process step, or of a failure, if the given process step could not be finished successfully.
	\item[Context Events] Events stemming from the context of the process, i.e., external sources, including IoT devices, sensors, third-party business partners collaborating in the process, and other data sources. These events are not directly connected to the process model, but can rather be correlated to the process steps.
\end{description}

We argue that the entirety of all events can be seen as an \emph{event stream}. A serialized form, either for transmission or for persistent storage, is an \emph{execution log} or \emph{event log}. Sometimes, the event log is part of the process log, e.g., \cite{Pika:2013}.

In our example, the event stream consists of both intrinsic events, i.e., the individual process steps like loading/unloading of containers, and context events, i.e., the readings from the sensor measuring the temperature of a banana container.
Note that a context event does not need to be related to a certain process step (i.e., to a series of intrinsic events), but may very well be used for failure prediction within multiple processes. 
For correlation, in our approach, it is assumed that the temporal co-presence of processes and the context event is sufficient.
Nonetheless, an explicit correlation can be also expressed \emph{a priori} so to limit the event scope and increase system scalability (e.g., detecting high temperature is relevant only during the delivery process, because it increases the probability of rotten bananas).

\subsection{Faults, Errors and Failures}

Differentiating the terms \emph{fault}, \emph{error} and \emph{failure} is significant in the context of failure prediction. In present literature, they have well-defined semantics \cite{avizienis04}: A \emph{fault} is the \emph{adjudged or hypothesized cause} of an error. An \emph{error} is the deviation of the system from its desired state. Finally, a \emph{failure} occurs when the system is not able to deliver its output as it is supposed to, leading to an undesirable outcome.

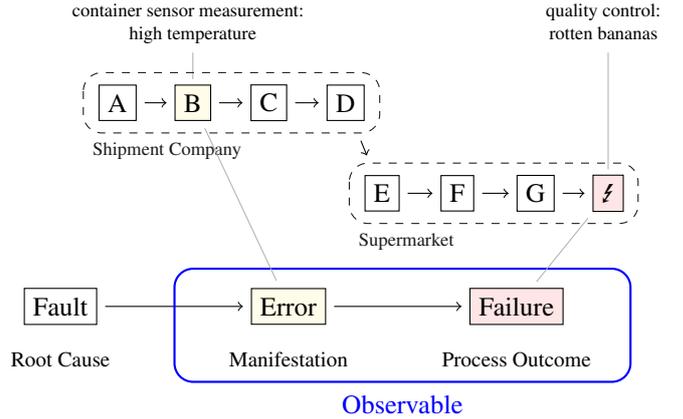
\begin{figure}
\begin{tikzpicture}

\tikzstyle{mystep}=[draw=black]
\tikzstyle{myw}=[anchor=west]
\tikzstyle{mydesc}=[]
\tikzstyle{mypath}=[shorten <=3pt, shorten >=3pt,->]
\tikzstyle{myconn}=[shorten <=3pt, shorten >=3pt, draw=black!30!white]
\tikzstyle{myconn2}=[shorten <=0pt, shorten >=2pt, draw=black!30!white]
\tikzstyle{partner}=[dashed, draw=black!90!white, rounded corners=7]
\tikzstyle{partnertext}=[anchor=north west, color=black!90!white]


\begin{scope}[shift={(0.5,0.5)}]
	\node[mystep, myw] (A) at (0,2.2) {A};
	\node[mystep, myw, fill=yellow!10!white] (B) at (1,2.2) {B};
	\node[mystep, myw] (C) at (2,2.2) {C};
	\node[mystep, myw] (D) at (3,2.2) {D};
	\node[mystep, myw] (E) at (3.5,1) {E};
	\node[mystep, myw] (F) at (4.5,1) {F};
	\node[mystep, myw] (G) at (5.5,1) {G};
	\node[mystep, myw, fill=red!10!white] (H) at (6.5,1) {\Lightning};

	\node[above of=B, node distance=.9cm] (Bx) {\scriptsize high temperature};
	\node[above of=Bx, node distance=.3cm] (Bx2) {\scriptsize container sensor measurement:\phantom{q}};

	\node[above of=H, node distance=2.1cm] (Hx) {\scriptsize rotten bananas\phantom{p}};
	\node[above of=Hx, node distance=.3cm] (Hx2) {\scriptsize quality control:\phantom{p}};

	\path[myconn2] (Bx) edge (B);
	\path[myconn2] (Hx) edge (H);

	\draw[partner] (-0.2,2.6) rectangle (3.7,1.8);
	\draw[partner] (3.3,1.4) rectangle (7.1,0.6);
	\path[mypath, shorten <=8pt, shorten >=8pt] (D) edge (E);
	\node[partnertext] at (-0.2,1.8) {\scriptsize Shipment Company};
	\node[partnertext] at (3.3,0.6) {\scriptsize Supermarket};
\end{scope}

\path[mypath] (A) edge (B);
\path[mypath] (B) edge (C);
\path[mypath] (C) edge (D);
\path[mypath] (E) edge (F);
\path[mypath] (F) edge (G);
\path[mypath] (G) edge (H);

\node[mystep] (fault) at (0,0) {Fault};
\node[mystep, fill=yellow!10!white] (error) at (3,0) {Error};
\node[mystep, fill=red!10!white] (failure) at (6,0) {Failure};

\path[mypath] (fault) edge (error);
\path[mypath] (error) edge (failure);

\node[mydesc, below of=fault, node distance=.7cm, align=center] {\footnotesize Root Cause};
\node[mydesc, below of=error, node distance=.7cm, align=center] {\footnotesize Manifestation};
\node[mydesc, below of=failure, node distance=.7cm, align=center] {\footnotesize Process Outcome};

\draw[thick, blue, rounded corners=7] (1.5, -1) rectangle (7.5, 0.5);
\node[blue] at (4.5, -1.3) {Observable};

\path[myconn] (B) edge (error);
\path[myconn] (H) edge (failure);

\end{tikzpicture}
\caption{Example of Faults, Errors and Failures as Parts of a Process.}
\label{fig:faults}
\end{figure}

In order to associate these semantics with our scenario, we define the application on these semantics to a BPMS. Figure~\ref{fig:faults} shows an example of a business process ending with a failure. In this example, some \emph{fault}, the root cause of the overall problem, has occurred, leading to an \emph{error} in the process. This error manifests itself during step B of the process, and can be detected by an observer. Finally, after steps C through G have passed, the overall failure of the system occurs.

It is noteworthy that time passes between the fault and the error, as well as between the error and the failure. Those delays may be very short, but they may also be very long, depending on the process. Since the fault itself is generally not observable, we can only detect its earliest manifestation, the error.

Without any prediction, the failure will only be visible upon its occurrence. However, adding a predictive element allows us to foresee a possible failure outcome at the first point in time it is detectable, i.e., when the first sign of an error occurs. In present literature, rule-based prediction is used to define the characteristics of errors~\cite{hermosillo10,zhao12}. In our proposed architecture, these rules are substituted by an ML component.

In the example presented earlier, a fault could consist of a defunct air cooling device, or an operator not activating it. The resulting error is a temperature exceeding a certain limit, and is detectable by a corresponding context event. Upon the reception of this event denoting excessive temperature, the failure, i.e., rotten bananas upon arrival, can be predicted.

\section{Solution Overview}
\label{sec:solution}

The integration of EBS and BPMS enables to control and adapt the execution of business processes at runtime by leveraging on intrinsic and context events.

During the execution of a business process in a BPMS, the concrete services instantiated for the tasks contained in the business process are executed, and this execution generates \textit{intrinsic events}. 
These events consist of task status changes, e.g., start and termination.
With respect to a specific business process, other and more detailed intrinsic events can be defined (e.g., delivery suspended, machine restarted, network connectivity unavailable).
Moreover, the service provider can enrich these events with Quality of Service (QoS) or non-functional information related to the service execution, such as the amount of resources or the monetary cost required to perform the task.

Furthermore, during its execution, a business process interacts with the environment (or context) that surrounds the invoked services and, in general, the BPMS. Therefore, we can identify external data sources, which, generating \textit{context events}, can enrich the process execution with further information. 
We argue that these context events must be correlated to the business process to a certain degree, either using temporal information, i.e., a sensor reading during the runtime of a process step, or using expert knowledge to define certain event sources as relevant for a certain step, i.e., defining the temperature of a container as relevant to the shipping step. Note that we do not necessarily need to have information about causal relationships, i.e., it is not necessary to define that an excessive temperature reading indicates an upcoming process failure. Instead, we merely define that the temperature sensor measurement is happening during the shipment step.
Although lots of data sources can be identified, the process of identifying the relevant ones, leading to the generation of valuable information, strictly depends on the specific business process and on the application that will exploit this data.
For example, if we want to monitor the shipping process and predict the ability to deliver on time, relevant data might come from weather monitors (e.g., to detect the presence of snow or heavy rain), route monitors (e.g., to predict traffic jams), or the presence of human agents who can slow down the process.

\begin{figure*}[!ht]
\centering
\includegraphics[width=0.7\textwidth]{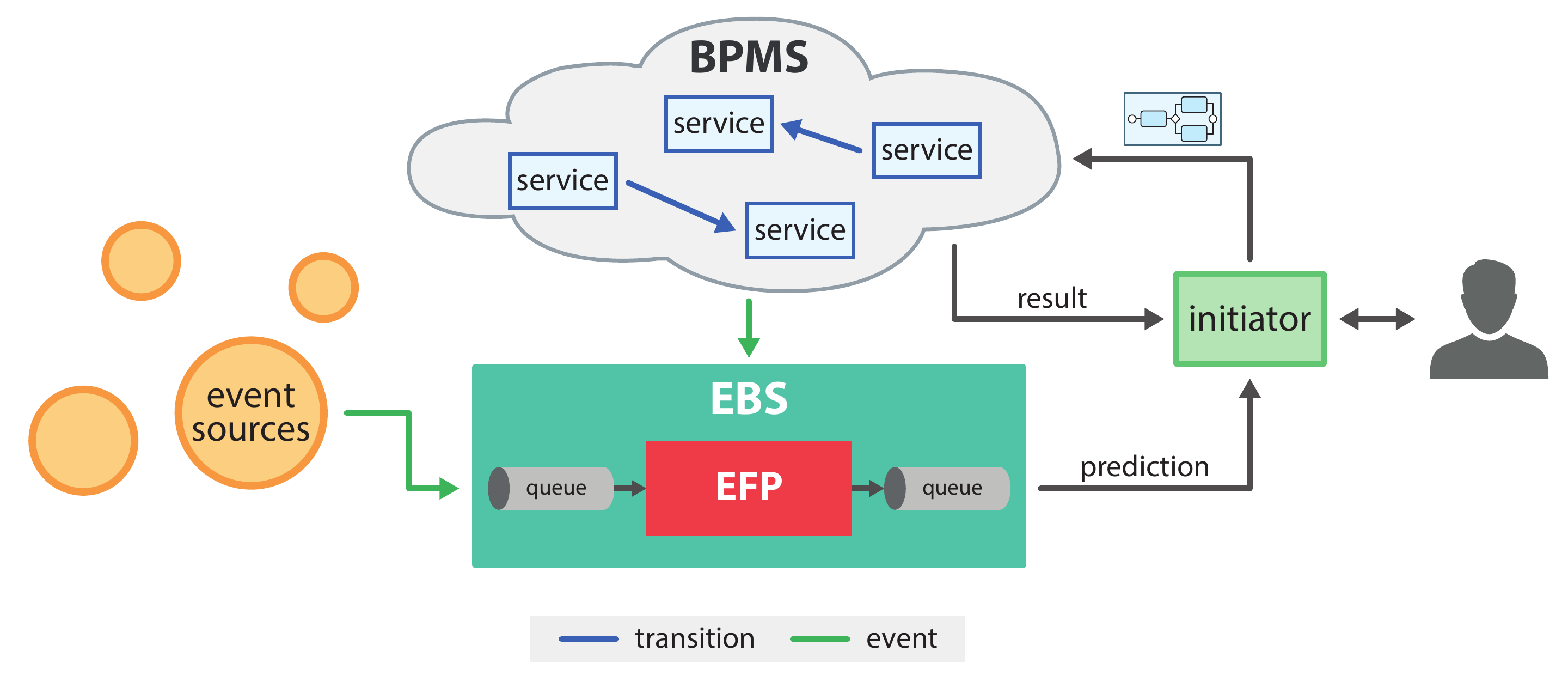}
\caption{Proposed System Architecture.}
\label{fig:overall-arch}
\end{figure*}

To show the benefits of the cooperation between BPMS and EBS, our solution exploits the huge amount of available data to 
perform an \textit{event-based failure prediction} (EFP) regarding the business process execution. 
Specifically, the EBS hosts and executes the EFP component which predicts the probability of failures at runtime, i.e., during the business process execution. 
The EFP component automatically learns the model of failure by leveraging on intrinsic and context events. 
Our solution is general enough to be able to predict failures related to functional and non-functional dimensions, i.e., it can identify an unsuccessful termination of the process (functional failure) or a termination with unsatisfying quality requirements, e.g., a product delivered with damages (non-functional failure).

A high-level representation of our solution is depicted in Figure~\ref{fig:overall-arch}. 
A user who wants to execute a business process interacts with an \textit{initiator} component which is in charge of two tasks: 
First, it launches the execution of the business process and the EFP component.
Second, it forwards to the user the result of the business process execution as well as the failure predictions emitted by the EFP component.
When the user asks for the execution of a business process, the initiator forwards its description, expressed as a workflow, to the BPMS. 
At the same time, the initiator triggers the EBS, which, in turn, creates a new instance of the EFP component that will predict failures for the newly started business process. Moreover, the initiator informs the external data sources of interest for the business process so that they can forward their context events to the EBS.
Acting as a surrogate of the user for the interactions with the other systems, the initiator can be highly distributed (not depicted in Figure~\ref{fig:overall-arch}).
Within the BPMS, each task of the business process is instantiated on a service that, aside performing its operations, generates intrinsic events. The amount and typology of data transported by these events depends on both the self-monitoring capability of the service and other motivations (e.g., security, privacy, political). 
Within the EBS, all the collected events are forwarded to the EFP component.

To reduce the coupling among the involved entities (i.e., EFP, services, event sources), the EBS uses a message queue system, where the distributed data sources publish intrinsic and context events. The EBS allocates a new queue for each business process execution. 
By subscribing to this queue, the EFP component receives, as a continuous stream, all the events related to the business process.
Since the EFP component is event-driven, each new event can trigger a failure prediction. To learn and identify failures, the EFP component exploits an online learning approach based on neural networks.
The details of the learning and prediction process will be presented in Section~\ref{sec:machine}.
As soon as a prediction is available, the EFP component publishes a new event on an outgoing queue, which is observed by several stakeholder services. Afterwards, each of these services can independently perform an operation, e.g., adapt the business process instance, notify the process owner, or trigger the execution of a different business process. As proof-of-concept of our approach, we implement the EFP component capable of processing an event stream stemming from a running business process, and of predicting possible imminent failures at runtime.

Scalability is a key point in EBS, which try to exploit parallelism to efficiently process incoming events (as our architecture does). 
Specifically, our architecture enforces separation of concerns and decouples the EBS from the BPMS by adopting a queuing service to distribute events. 
We also devise EFP to control one business process at a time; as such, multiple processes can be regarded by multiple independent EFP components. 
Albeit challenging, scaling the EFP component is in line with the current research trends~\cite{bello2016social, li2014scaling}, which propose, for example, approaches to achieve distributed ML capabilities~\cite{kraska2013mlbase, meng2016mllib}.
Empirical evidence~\cite{Mayer:2017} shows how ML approaches can efficiently deal with big amounts of data in (near) real-time fashion.

So far, we have considered business processes in a generic manner, without accounting for the organizations involved in their execution. 
As a specific case, apart from intra-organization business processes, we can distinguish  inter-organizational ones, which involve the collaboration among multiple partners for their fulfillment. 
The presence of multiple parties brings up the issue of privacy. Indeed, some organizations might not share information about their private processes, thus limiting the visibility of their private events.
In Section~\ref{sec:evaluation}, we explicitly consider this critical point while evaluating the efficacy of the proposed approach.

\section{Machine Learning Failure Prediction}
\label{sec:machine}
The basic idea of our approach is the assumption that process execution failures can be predicted based on events, and that therefore it is possible to identify early deviations from the expected process behavior. In theory, rules could be put in place to identify such deviations and react accordingly, whenever necessary. However, the definition of rules has certain drawbacks:

\begin{itemize}
	\item Creating rules is a task requiring expert (domain-specific) knowledge and time, as expertise on the relationship between certain events is needed to formulate rules.
	\item Ensuring exhaustion of all possible rules is difficult. If the expert is unaware of a certain relationship between an event and the outcome of a process step, the missing rule goes unnoticed.
	\item Concept drift \cite{widmer1996learning} is affecting manually created rules. If relationships between events and failures change, existing rules might become obsolete or wrong, and lead to mispredictions (false positives or false negatives). This can only be overcome by periodic reviews, which in turn require expert knowledge and time.
\end{itemize}

In order to overcome these issues, we propose to use methods from the field of ML in order to automate the process of predicting failures. Not only does this approach require substantially less expert knowledge and therefore facilitates solutions with increased abstraction from the domain, but it also promises to find relationships not yet discovered. Furthermore, concept drift can be mitigated by methods well-established in ML \cite{widmer1996learning,klinkenberg2000detecting}.

In fact, the solution presented in this paper can co-exist with expert-generated rules. For instance, present rules can support the initial training phase, during which the ML model might not produce meaningful output. After initial training, the ML model can be used to subsequently verify whether present rules are still valid, or have been made obsolete by concept drift.

\subsection{Failure Prediction Component}

The EFP component consists of two interacting systems. The first system is responsible for the prediction itself: While a process is executed by the BPMS, and events are populated through the EBS, the EFP component is analyzing these events. Based on this analysis, the EFP component might indicate that with a certain probability, a failure is going to occur in a given future process step (or at the end of the process). Should such a situation arise, the EFP component fires an event to notify event queue subscribers, which can then prompt the user and evaluate possible steps to mitigate the risk of a failure occurring. The second subsystem is a training component. After each completed process execution, a trace of the recorded events (including all executed steps) is used to train the prediction model, increasing accuracy in subsequent runs. Nevertheless, since the model is performing its predictions on an event stream, and does not need to store the entire data set in-memory (only the event trace of the current process execution), we classify our methodology as an online learning approach.

At the core of our solution, a specially designed artificial neural network (ANN) model is responsible for analyzing the stream of incoming data in order to perform a failure prediction.
ANN models consist of interconnected layers of perceptrons, each of which is aggregating input data, processing this data using a so-called \emph{activation function}, and passing the output either to the next layer of neurons, or to the network output in case of the last layer \cite{neural}.

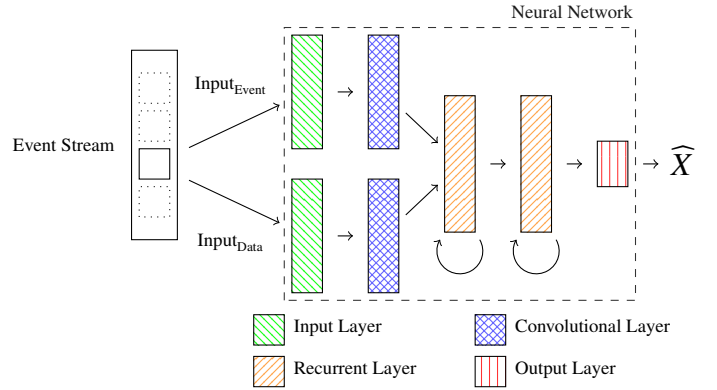
\begin{figure}[ht]
	\centering
	\begin{tikzpicture}

\tikzstyle{mystep}=[draw=black]
\tikzstyle{myw}=[anchor=west]
\tikzstyle{mydesc}=[]
\tikzstyle{mypath}=[shorten <=3pt, shorten >=3pt,->]
\tikzstyle{myconn}=[shorten <=3pt, shorten >=3pt, draw=black!30!white]


\begin{scope}[shift={(-3.0,0)}]
	\draw (-0.30,-1.0) rectangle (0.30,1.5);

	\draw[dotted] (-0.2, -0.7) rectangle (0.2, -0.3);
	\draw (-0.2, -0.2) rectangle (0.2, 0.2);
	\draw[dotted] (-0.2, 0.3) rectangle (0.2, 0.7);
	\draw[dotted] (-0.2, 0.8) rectangle (0.2, 1.2);

\node[anchor=east] at (-0.4, 0.25) {\scriptsize Event Stream};
\end{scope}

\draw[shorten <=0.5cm, shorten >=0.4cm, ->] (-3.0,0) -- (-1,0.95);
\draw[shorten <=0.5cm, shorten >=0.4cm, ->] (-3.0,0) -- (-1,-0.95);
\node[anchor=south] at (-2.0, 0.8) {\scriptsize $\text{Input}_\text{Event}$};
\node[anchor=north] at (-2.0, -0.8) {\scriptsize $\text{Input}_\text{Data}$};

\begin{scope}[shift={(-1.0,0)}]
	\draw[pattern=north west lines,pattern color=green!90!white] (-0.2, 1.7) rectangle (0.2, 0.2);
	\draw[pattern=north west lines,pattern color=green!90!white] (-0.2, -1.7) rectangle (0.2, -0.2);
\end{scope}

\draw[shorten <=0.4cm, shorten >=0.4cm, ->] (-1,0.95) -- (0.0,0.95);
\draw[shorten <=0.4cm, shorten >=0.4cm, ->] (-1,-0.95) -- (0.0,-0.95);

\begin{scope}[shift={(0.0,0)}]
	\draw[pattern=crosshatch,pattern color=blue!60!white] (-0.2, 1.7) rectangle (0.2, 0.2);
	\draw[pattern=crosshatch,pattern color=blue!60!white] (-0.2, -1.7) rectangle (0.2, -0.2);
\end{scope}

\draw[shorten <=0.4cm, shorten >=0.4cm, ->] (0.0,0.95) -- (1.0,0.0);
\draw[shorten <=0.4cm, shorten >=0.4cm, ->] (0.0,-0.95) -- (1.0,0.0);

\begin{scope}[shift={(1.0,0)}]
	\draw[pattern=north east lines,pattern color=orange!70!white] (-0.2, -0.9) rectangle (0.2, 0.9);
	\begin{scope}[shift={(0,-1.15)}]
		\draw[->] (30:0.3cm) arc (30:-210:0.3cm);
	\end{scope}
\end{scope}

\draw[shorten <=0.4cm, shorten >=0.4cm, ->] (1.0,0.0) -- (2.0,0.0);

\begin{scope}[shift={(2.0,0)}]
	\draw[pattern=north east lines,pattern color=orange!70!white] (-0.2, -0.9) rectangle (0.2, 0.9);
	\begin{scope}[shift={(0,-1.15)}]
		\draw[->] (30:0.3cm) arc (30:-210:0.3cm);
	\end{scope}
\end{scope}

\draw[shorten <=0.4cm, shorten >=0.4cm, ->] (2.0,0.0) -- (3.0,0.0);

\begin{scope}[shift={(3.0,0)}]
	\draw[pattern=vertical lines,pattern color=red!80!white] (-0.2, -0.3) rectangle (0.2, 0.3);
\end{scope}

\draw[shorten <=0.4cm, shorten >=0.4cm, ->] (3.0,0.0) -- (4.0,0.0);

\node at (3.9,0.05) {\large $\widehat{X}$};

\begin{scope}[shift={(-1.5,-1.9)}]
	\begin{scope}[shift={(0.0,0.0)}]
		\draw[pattern=north west lines,pattern color=green!90!white] (-0.2, -0.09) rectangle (0.2, -0.49);
		\node[anchor=north west] at (0.2,-0.03) {\scriptsize Input Layer};
	\end{scope}
	\begin{scope}[shift={(2.9,0.0)}]
		\draw[pattern=crosshatch,pattern color=blue!60!white] (-0.2, -0.09) rectangle (0.2, -0.49);
		\node[anchor=north west] at (0.2,-0.03) {\scriptsize Convolutional Layer};
	\end{scope}
	\begin{scope}[shift={(0.0,-0.55)},pattern color=orange!70!white]
		\draw[pattern=north east lines] (-0.2, -0.09) rectangle (0.2, -0.49);
		\node[anchor=north west] at (0.2,-0.03) {\scriptsize Recurrent Layer};
	\end{scope}
	\begin{scope}[shift={(2.9,-0.55)},pattern color=red!80!white]
		\draw[pattern=vertical lines] (-0.2, -0.09) rectangle (0.2, -0.49);
		\node[anchor=north west] at (0.2,-0.03) {\scriptsize Output Layer};
	\end{scope}
\end{scope}

\draw[draw=black!80!white, dashed] (-1.3,-1.8) rectangle (3.3,1.8);
\node[color=black!90!white, anchor=south east] at (3.39,1.8) {\scriptsize Neural Network};

\end{tikzpicture}
	\caption{Schematic Representation of the Proposed ANN Model.}
	\label{fig:network}
\end{figure}

As it can be seen in Figure~\ref{fig:network}, our network consists of both convolutional and recurrent layers. Convolutional layers are layers of perceptrons aggregating input from neurons which are semantically similar, and have proven useful, e.g., for facial recognition \cite{facial} or natural language processing~\cite{language}. In our case, this aids to reduce the network's sensibility to temporal sequences of events. Furthermore, the convolutional layers help us to use different kinds of input in one ANN. Recurrent layers introduce circles into otherwise acyclical graphs of perceptrons; in our case, we use Long Short-Term Memory (LSTM) layers~\cite{lstm}. This adds state information to the network, which is used to process temporal sequences of events. The combination of both convolutional and recurrent layers combines the power of both to support the training of temporal data while avoiding excessive sensibility of the ANN.

For selecting the activation functions of our ANN, we have followed work from \cite{lecun2015deep}, selecting the \emph{rectifier} as the main activation function. This function has the advantage of a lower bound (\emph{cutoff}), below which no activation occurs. It has been presented in \cite{hahnloser2000digital} and is, due to its effectiveness, an activation function commonly used in recent years \cite{lecun2015deep}. In contrast, both the $tanh$ function and the very similar \emph{sigmoid} (logistic) curve approach their lower bound but only reach it in infinity, i.e., some residual activation occurs, no matter how low the sum of weighted inputs is. Due to the fact that our problem is a classification problem, it has a highly discrete nature, and is therefore profiting from the cutoff of the rectifier function. However, initial experiments have shown that the first layer, i.e., the first convolutional layer after the input layer, performed best using the $tanh$ activation function.

Compared to other ML models like decision trees, support vector machines etc., ANNs have the disadvantage of being rather black-boxed models. First, constructing an ANN and deciding on its topology is a non-trivial task, often involving a lot of exploration and experimentation. Second, the resulting rules are not human-readable rules, but trained weights, resulting in a mathematical model without clear semantics for the single neurons. Nevertheless, neural networks, together with recurrent layers and convolutional layers, are well-suited for processing temporal data such as time series or event logs, maintaining an internal state \cite{neural}.

\subsection{Input and Output Structure}

As described before, the ANN model is presented with input data both during training as well as during the actual prediction phase. In the case of training, we also provide output data (labels) to the network for supervised learning. The input data consists of the type of event captured as well as the data associated with the event, if any. As described in Section~\ref{eventstreams}, we distinguish between intrinsic events, stemming from the process itself, and context events, stemming from the external context of the process.

In our notation, we denote the types of process-intrinsic events (i.e., the possible process steps) as $I_\text{fail}, I_0, I_1, \dots, I_n$, where $I_\text{fail}$ represents a failure in the current step, and the remaining symbols $I_0, \dots, I_n$ represent all possible process steps. Context events, e.g., generated by sensors, are denoted as $C_0, C_1, \dots, C_m$.

We therefore define the input vector for the ANN as follows:
\begin{align}
\text{Input} &=  [ \text{Input}_\text{Event}, \text{Input}_\text{Data}] \\
\text{Input}_\text{Event} &= [ I_\text{fail}, I_0, I_1, \dots, I_n, C_0, C_1, \dots, C_m ] \\
\text{Input}_\text{Data} &= [ D_0, D_1, \dots, D_k ]
\end{align}
where $n$ and $m$ are the number of intrinsic and context events, respectively, known to the system. $\text{Input}_\text{Event}$ is a binary vector consisting of one variable $I_i$ for each intrinsic event type in the process and one variable $C_i$ for each context event type. Depending on the type of the incoming event, either exactly one variable $I_i$ is $1$ for the intrinsic event $I_i$, or exactly one variable $C_i$ is $1$ for the context event type $C_i$ in a given input row; all other variables are $0$. Furthermore, $\text{Input}_\text{Data}$ is a vector containing the data associated with the event, if any. This data might include, for instance, the sensor reading from a temperature sensor. The cardinality of $\text{Input}_\text{Data}$ ($k$) depends on the type of event, i.e., which one of the variables $I_\text{fail}, I_0, I_1, \dots, I_n, C_0, C_1, \dots, C_m$ is $1$.

The output of the ML model is a classification of what the next step of the model will be (or whether it will be a failure). The process steps contained in the process model correspond to $I_\text{fail}, I_0, I_1, \dots, I_n$, therefore, the output is a vector giving, for each process step, the probability that this process step will be the next one. Note that for the sake of simplicity, we only regard one execution branch of a collaboration or choreography process. However, this does not limit the applicability of the proposed model to multiple processes running in parallel. Following this, we formulate the following output vector structure:
\begin{equation}
\text{Output} = [ \widehat{I_\text{fail}}, \widehat{I_0}, \widehat{I_1}, \dots, \widehat{I_n} ]
\label{eq:output}
\end{equation}

\subsection{Formalization Model}

In order to formally describe the underlying problem and our approach, we introduce a model for reasoning about the predictions of process outcomes. To this end, we build upon the model of probabilistic automata (PAs) \cite{RABIN1963,salomaa2014theory}. A PA is a generalization of a non-deterministic finite automaton (NFA), where instead of a membership function determining which states can be reached with which input, these binary values are substituted by probabilities. Formally, a PA consists of the following attributes \cite{RABIN1963}:

\begin{itemize}
	\item A finite set of states $Q$.
	\item A finite set of input symbols $\Sigma$.
	\item A transition matrix $P$.
	\item An initial state\footnote{Note that some literature uses a \emph{distribution vector} for determining the initial state~\cite{salomaa2014theory} instead of a fixed single state $q_0$. We use a single initial state, since the process model can be assumed to have a fixed initial state, and this simplifies the definition without reducing expressive power.} $q_0 \in \Sigma$.
	\item A set of states $F \subset Q$ which are defined as final states.
\end{itemize}

In our model, the set of states $Q$ is the set of process steps, including the \emph{failure} state $q_\text{fail}$, representing a failure in a business process. For the set of input symbols $\Sigma$, we use the set of input and context events read by the predictor. Our initial state $q_0$ is the start state of the business process. The set of final states $F$ is equal to the set of end states of the business process at hand. By definition, the failure state is also a final state, i.e., $q_\text{fail} \in F$.

The transition (stochastic or probability) matrix $P$ determines the probability for the automaton to enter another state, given a current state and an input. A common notation in the definition of PA is $p_j(q_i, x) \in P$, denoting the probability for the PA, with the current state $q_i$ and given the input $x$ to enter state $q_j$ \cite{salomaa2014theory}. Naturally, the sum of the probabilities for all subsequent states of a state $q_i$ and an event $s$ is 1, since it is certain that \emph{some} state must be reached:
\begin{equation}
	\sum_{j} p_j(q_i, x) = 1
\end{equation}

In our model, the probability matrix $P$ is not a fixed matrix. Instead, whenever a prediction of the following step is required, the previously described ANN is invoked, yielding probability values for all possible next steps. In other words, a row of the transition matrix is returned, as seen in (\ref{eq:output}).

We now define that, at any point in time during the execution of a process, there is an event trace $T$, which consists of all recorded events (including intrinsic events, i.e., state changes, and context events). We argue that the current process step is deducible from this event trace by merely searching for the last recorded intrinsic event indicating a process step, and define that step as $q_i$. In order to predict the subsequent process steps to deduce whether a failure might occur, we are interested in the probabilities for the process to continue with a certain step $q_{i+1}$. As discussed, instead of a fixed transition matrix to determine the probable next step $q_{i+1}$, we use the ANN which returns, for each possible step $q \in Q$, the probability for this step. We denote the application of the ANN model onto a given step trace~$T$ as $\widehat{X}(T)$.

\subsection{Probability Traversal}

Following the previous definitions, we describe the further processing of predictions by the predictor component. We have already defined an event trace $T$, which denotes the already-recorded (historic) events and stems from a running process instance. We invoke the previously discussed ANN onto $T$, yielding a row vector out of the transition matrix; we call this vector $P_T$:
\begin{equation}
	P_T = \widehat{X}(T)
\end{equation}

For each step $q \in Q$ (corresponding to the states of the PA), $P_T(q)$ yields the probability of the process to continue with this step $q$. This probability for a single step is called \emph{step} probability. The sum of all step probabilities for a given event trace $T$ is 1, since \emph{some} step, possibly $q_\text{fail}$, is certainly going to be the subsequent one.

\newcommand{\PP}{\ensuremath{\mathbb{P \hspace{0.1em}}}}
\newcommand{\OO}{\ensuremath{\Omega \hspace{0.1em}}}

\begin{figure}[t]
	\centering
	\begin{tikzpicture}

\tikzstyle{mystep}=[draw=black, minimum width=0.55cm, anchor=west]
\tikzstyle{myp}=[minimum width=0.55cm, anchor=west]

\tikzstyle{myw}=[anchor=west]
\tikzstyle{mydesc}=[]
\tikzstyle{mypath}=[shorten <=3pt, shorten >=3pt,->]
\tikzstyle{myconn}=[shorten <=3pt, shorten >=3pt, draw=black!30!white]


\newcommand{\pr}[1]{#1}

\node[mystep] (A) at (0.0,0.0) {A};
\node[mystep] (B) at (1.5,0.0) {B};
\node[mystep] (C) at (3.0,0.0) {C};

\node[mystep] (D) at (4.5,1.0) {D};
\node[myp] (Dp) at (5.15,1.0) {0.803};
\node[mystep] (E) at (4.5,0.0) {E};
\node[myp] (Ep) at (5.15,0.0) {0.010};
\node[mystep] (f1) at (4.5,-1.0) {\Lightning};
\node[myp] (fp) at (5.15,-1.0) {0.187};

\node[mystep] (F) at (6.9,1.5) {G};
\node[myp] (Fp) at (7.55,1.5) {0.005};
\node[mystep] (f2) at (6.9,0.5) {\Lightning};
\node[myp] (Fp) at (7.55,0.5) {0.995};

\draw[<-,color=red] (6.4,-1.0) -- (7.0,-1.0);
\node[anchor=west,color=red] at (7.0,-1.0) {\scriptsize failure};
\draw[<-,color=red] (8.9,0.5) -- (9.4,0.5);
\node[anchor=west,color=red] at (9.4,0.5) {\scriptsize failure};

\draw[mypath] (A) -- (B);
\draw[mypath] (B) -- (C);
\draw[mypath] (C) -- (D);
\draw[mypath] (C) -- (E);
\draw[mypath] (C) -- (f1);
\draw[mypath] (Dp) -- (F);
\draw[mypath] (Dp) -- (f2);

\draw[->] (-0.5,-2.0) -- (9.5,-2.0);
\node at (9.5,-2.3) {time};
\draw (4.0,-1.9) -- (4.0,-2.1);

\draw[fill=black!5!white] (0,-1.8) rectangle (3.8, -2.2);
\node at (1.9,-2.0) {\scriptsize history (past)};

\draw[fill=black!5!white] (4.2,-1.8) rectangle (9.0, -2.2);
\node at (6.6,-2.0) {\scriptsize prediction (future)};

\draw[color=black!80!white,decorate,decoration={brace,amplitude=10pt,raise=4pt},yshift=0pt] (0.0,2.1) -- (3.55,2.1);
\node[color=black!80!white] at (1.75,3.0) {\scriptsize execution trace $T$};

\node[color=black!80!white] at (5.8,3.31) {\scriptsize probability};
\node[color=black!80!white] at (5.8,3.0) {\scriptsize $P_T(\cdot)$};
\draw[color=black!80!white,->] (5.8,2.7) -- (5.8,1.5);

\node[color=black!80!white] at (8.2,3.31) {\scriptsize probability};
\node[color=black!80!white] at (8.2,3.0) {\scriptsize $P_{T \rightarrow D}(x)$};
\draw[color=black!80!white,->] (8.2,2.7) -- (8.2,2.0);

\end{tikzpicture}
	\caption{Example Tree for the Trace $A \ra B \ra C$, Showing Possible Subsequent Events, Including Failures, and their Step Probability. Elements E and G are Defined as Final States.}
	\label{fig:tree}
\end{figure}
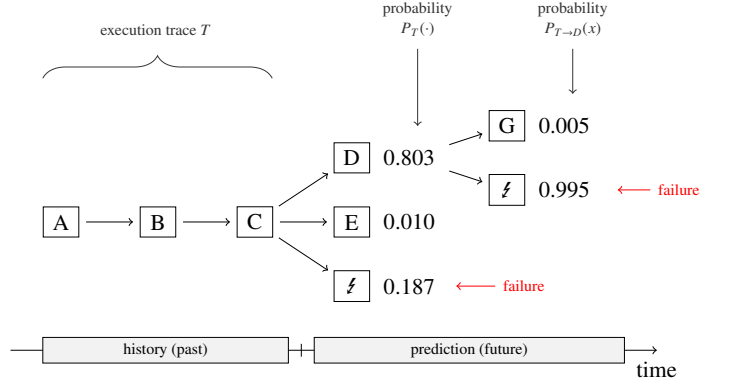

For instance, Figure~\ref{fig:tree} shows a trace of events $T$, containing the events $A \rightarrow B \rightarrow C$. The probability vector $P_T$ has been used by invoking the ANN classifier $P_T = \widehat{X}(T)$. The values for $P_T$ are shown in Table~\ref{fig:table1}.

\begin{table}[h!]
	\centering
	\caption{Values of the Probability Vector $P_T$, for all Elements with Non-zero Probability, with $T = A \rightarrow B \rightarrow C$, Corresponding to the Example Shown in Figure~\ref{fig:tree}.}
	\begin{tabular}{ l c c }
		$q$ & $P_T(q)$ \\
		\hline
		D & 0.803 \\
		E & 0.010 \\
		$q_\text{fail}$ & 0.187 \\
		\hline
		$\Sigma$ & 1.000
	\end{tabular}
	\label{fig:table1}
\end{table}

Similarly, this traversal is also performed for subsequent steps. For instance, Table~\ref{fig:table2} shows the resulting probability vector for the trace $T \ra D$. %

\begin{table}[h!]
	\centering
	\caption{Values of the Probability Vector $P_{T \ra T'}$, for all Elements with Non-zero Probability, with $T = A \rightarrow B \rightarrow C$ and $T' = D$, Corresponding to the Example Shown in Figure~\ref{fig:tree}.}
	\begin{tabular}{ l c c }
		$q$ & $P_{T \ra T'}(q)$ \\
		\hline
		G & 0.005 \\
		$q_\text{fail}$ & 0.995 \\
		\hline
		$\Sigma$ & 1.000
	\end{tabular}
	\label{fig:table2}
\end{table}

In this manner, we traverse the entire space of possible subsequent steps, and for each possible step $q$, re-evaluate all following possible steps. This traversal is done until an end step is reached, at which point the traversed trace is recorded, together with its total probability and its outcome.

The total probability is the conditional probability of a given trace $T$ to be followed by the future step sequence $T'$, and is denoted as $\PP(T \rightarrow T')$. The outcome defined as $\OO(T')$ denotes how the sequence $T'$ ends, and is either \emph{end}, for an orderly finished process, or \emph{fail}, if a failure occurred, i.e., if $q_\text{fail}$ was reached. The probability \PP is defined as follows:
\begin{align}
	\PP(T) &= 1 \label{pp1} \\
	\PP(T \ra T') &= \PP(T) \cdot \PP(T') \label{pp2} \\
	\PP(T' \ra q) &= \PP(T') \cdot \PP(q) \label{pp3} \\
	\PP(q) &= P_T(q) \label{pp4}
\end{align}

In (\ref{pp1}), we define the total probability of the original event trace $T$ as 1, since the event trace has already been recorded, and thus its occurrence is certain. In (\ref{pp2}), the total probability of the event trace $T$ followed by an event sequence $T'$ is defined as the product of the total of probability of $T$ and the (partial) total probability of $T'$. Finally, (\ref{pp3}) and (\ref{pp4}) define that the partial total probability of $T'$ is the product of its elements: (\ref{pp3}) defines that the partial total probability of a sequence followed by an element is the product of the respective probability, and (\ref{pp4}) defines that the partial total probability of a single element is its step probability.

Naturally, the sum of the total probabilities of all possible event sequences following a given trace $T$ is 1, since \emph{some} event sequence, possibly one where the outcome is a failure, will eventually be the resulting sequence of the process.

To formalize our traversal, we define the \textit{traverse} function of an event sequence $T$, which aggregates the results yielded by $\widehat{X}(\cdot)$:
\begin{equation}
\textit{traverse}(T) = \bigcup_{q \in Q} \textit{visit}(T \ra q)
\end{equation}
where $\textit{visit}(\cdot)$ is the function generating a set of resulting sequences, based on this event sequence $T \ra q$. The function $\textit{visit}(\cdot)$ is defined as follows:
\begin{align}
\textit{visit}(T \ra q) &= 
	\begin{cases}
		T \ra q, & \text{if $q \in F$}\\
		\textit{traverse}(T \ra q) & \text{otherwise}
	\end{cases}
\end{align}

As we can see, the \textit{visit} function, upon encountering a non-final element $q$, invokes the \textit{traverse} function again using the concatenation of $T$ and $q$, i.e., $T \ra q$. This makes \textit{traverse} a recursive function. Table~\ref{fig:table} shows the resulting sequences of the example process, together with their probabilities and outcomes.

\begin{table}[t!]
	\centering
	\caption{Sequence, Probabilities and Outcomes Resulting from the Tree in Figure~\ref{fig:tree}, with $T = A \ra B \ra C$.}
	\begin{tabular}{ l c c }
		T' & $\PP(T \ra T')$ & \OO(T') \\
		\hline
		$D \ra q_\text{fail}$ & 0.799 & \emph{fail} \\
		$q_\text{fail}$ & 0.187 & \emph{fail} \\
		E & 0.010 & \emph{end} \\
		$D \ra G$ & 0.004 & \emph{end} \\
		\hline
		$\Sigma$ & 1.000 &
	\end{tabular}
	\label{fig:table}
\end{table}

\begin{figure*}[t]
\removelatexerror
\begin{algorithm}[H]
{ \small
	\Fn{Traverse ($T$)} {
		\KwData{$T$ contains the event trace from the currently running process}
		\KwResult{List of all possible process outcomes, starting from the last captured event, along with their probability values}
		\BlankLine
		\Return Recurse($T$, 1.0)\;
	}
	\BlankLine

	\Fn{Recurse ($T$, $P_\text{curr}$)} {
		\KwData{$T$ contains an event trace $T = [T_0, \dots, T_i]$ that is assumed to be already fixed}
		\KwData{$P_\text{curr}$ is the total probability of the current event trace}
		\KwResult{List of all possible further process traces, starting from $T_i$, along with their probabilities}
		\BlankLine

		\lIf(\tcp*[f]{Depth Limit}){$|T| > \texttt{MAX\_DEPTH}$}{\Return []} 
		\lIf(\tcp*[f]{Probability Limit}){$P_\text{curr} < \texttt{MIN\_PROBABILITY}$}{\Return []}
		\BlankLine

		Result $\leftarrow []$\;
		Breadth $\leftarrow 0$\;
		\BlankLine
		\tcp{Fetch predictions from ANN classifier by feeding it to the event trace}
		Pred $\leftarrow$ GetPredictionsFromClassifier($T$)\;
		\emph{sort} Pred \emph{by probability descending}\;
		\BlankLine
		\ForEach{$e \in \text{\emph{Pred}}$}{
			NextStep $\leftarrow e$\;
			NextStepProbability $\leftarrow$ Pred[$e$]\;
			\BlankLine

			\lIf(\tcp*[f]{Breadth Limit}){$\text{++Breadth} > \texttt{MAX\_BREADTH}$}{\Continue}
			\BlankLine

			\uIf{$e$ is end state}{
				Trace $\leftarrow T \oplus e$ \tcp*{$\oplus$ is the concatenation operator}
				Probability $\leftarrow P_\text{curr} \cdot \text{NextStepProbability}$\;
				Outcome $\leftarrow$ End state $e$ reached\;
				\emph{add} $<\text{Trace}, \text{Probability}, \text{Outcome}>$ \emph{to} Result\;
			}
			\uElseIf{$e$ is failure}{
				Trace $\leftarrow T \oplus \text{failure}$ \tcp*{$\oplus$ is the concatenation operator}
				Probability $\leftarrow P_\text{curr} \cdot \text{NextStepProbability}$\;
				Outcome $\leftarrow$ Failure in $e$\;
				\emph{add} $<\text{Trace}, \text{Probability}, \text{Outcome}>$ \emph{to} Result\;
			}
			\Else{
				NextPossibilities $\leftarrow$ Recurse($[T, e]$, $P_\text{curr} \cdot \text{NextStepProbability}$)\;
				\emph{add} NextPossibilities \emph{to} Result\;
			}
		}

		\Return Result\;
	}
	}
\caption{Algorithm for Bounded Traversal.}
\label{alg:traverse}
\end{algorithm}
\end{figure*}

Putting the definitions together, the predictor component can, at any given point in time during the execution of a process, use the trace of already-recorded events $T$, and by invoking $\textit{traverse}(T)$, build a list of possible future event sequences, along with their probabilities \PP{} and outcomes \OO.

\subsection{Search Space Optimization}

For large process models, the search space defined by the recursive function \textit{traverse} can become too large to be processed in an \emph{online} matter, i.e., during the process execution. If the processing time increases, the outcomes might not be predicted in time. To mitigate this, we propose several ways of limiting the search space of the recursive algorithm.

\begin{description}
	\item[Process Model Correlation] Since the underlying predictor uses an ANN model to predict subsequent events (including steps) in a process, the result of this prediction might include events which are not possible in the current state. For instance, if there is no control flow relation between steps C and D in the example shown in Figure~\ref{fig:tree}, and the last event in the recorded event sequence is C, D is not a possible subsequent event, and this part of the tree does not need to be explored. With a na\"{\i}ve search, however, the classifier could predict this impossible sequence, and it could even be predicted as the \emph{most likely} sequence. This is especially the case during the phase of initial training, or in unusual or novel event sequences. The reason for this behavior is the fact that the classifier itself does not take into account the process model being executed. Therefore, we introduce a stopping condition for the traversal. This condition filters out event combinations that are not possible according to the underlying process model.

	\item[Probability Limit] In a similar manner as with the previously discussed event sequences not possible according to the process model, we also disregard highly unlikely events. Our solution introduces a probability parameter \texttt{MIN\_PROBABILITY} which is applied to the total probability of the entire path. In other words, a possible event branch is not traversed further if its (partial) total probability is below a threshold.

	\item[Depth and Breadth Limit] In addition to a minimum probability, we introduce the parameters \texttt{MAX\_DEPTH} and \texttt{MAX\_BREADTH} defining the upper bound for depth and breadth within our search. Limiting the search depth is based on the fact that predicting events which are too distant in the future may become decreasingly meaningful. Limiting the search depth numerically is working together with limiting the probability to reduce the search space.
\end{description}

The limits represent hyper-parameters of our solution and are used to maintain a certain upper bound on the traversal runtime. The primary goal of this bound is to avoid infinite traversal in processes with cycles (e.g., loops).

Algorithm~\ref{alg:traverse} shows a consolidated, algorithmic form of all the calculations described above. While the main function, \textit{Recurse}, contains the main (recursive) logic, the first function, \textit{Traverse}, shows the initial call for \textit{Recurse}. The \textit{Recurse} function takes two parameters: $T$, which is the process trace which is assumed as already known, and $P_\textit{curr}$, which determines the total probability until the current step, as defined in (\ref{pp1})-(\ref{pp4}). Naturally, \textit{Traverse} initializes this total probability with 1, because the actually recorded history of steps has already happened, and therefore has a probability of 1, as defined in (\ref{pp1}). Lines 4 and 5 represent the depth and probability limits, respectively. Lines 6 and 7 initialize local variables. Line 9 calls the ANN in order to obtain a vector of probability values for each possible subsequent step. This vector is sorted in line~10 by probability.

The loop from line 11 to line 29 iterates over the $n$ most likely subsequent steps, where $n = \texttt{MAX\_BREADTH}$ (ensured by line 14, the breadth limit), and checks whether they are an end state, a failed state, or a normal process step. For end states, lines 16-19 add an element to the result vector. Similarly, lines 21-24 do the same for a failure. 

\section{Evaluation}
\label{sec:evaluation}
This section presents an empirical assessment of the proposed approach following the \emph{single-case mechanism experiments} validation method \cite{DBLP:books/sp/Wieringa14}. The evaluation was conducted on two different data sets: (i) a real-world data set from the finance domain, and (ii) a simulated data set of a realistic distributed manufacturing process. Both data sets were pre-processed and used to train and assess the performance of the ML models in detecting failures. The conducted experiments prove the applicability and feasibility of combining EBS and distributed business processes in a real-world scenario.  

Section \ref{sec:sources} discusses the selection of the two data sets used for our evaluation. Afterwards, Sections~\ref{sec:prep-real} and~\ref{sec:prep-synth} show how the data sets were pre-processed, and Section~\ref{sec:faults} describes the mechanisms necessary for enriching the data sets with faults and failures. Finally, Section~\ref{sec:experiments} presents the experiments performed and gives an overview of the results.

\subsection{Data Sets}
In order to conduct experiments to evaluate the feasibility of our approach, we explored various data sets of different domains, and studied their usability and suitability for the evaluation of our approach.

\subsubsection{Explored Sources of Data Sets}
\label{sec:sources}
The search for an appropriate data set for the experiments was conducted with respect to the following criteria: Whether the data set is (i)~publicly available to the research community, (ii)~event-based, (iii)~correlated with business process models (i.e., events are correlated with process tasks and instances), (iv)~well-documented, (v)~contains context events and (vi)~failures, and (vii)~stems from an inter-organizational process. Note that unless a data set that satisfies all the criteria is found, a compromise over the criteria must be considered for the data set selection. 
 
A multitude of data sources have been examined from either projects, research challenges or other public data sources. In particular, data sets from the BPI Challenge\footnote{\url{https://www.win.tue.nl/bpi/doku.php?id=2017:challenge}} and the Kaggle Competition\footnote{\url{https://www.kaggle.com/}} were taken into account. 
Platforms for competitions in the domain of data science and ML, such as Kaggle, are naturally a valuable source for test data. Public directories for data sets are also available\footnote{\url{https://www.springboard.com/blog/free-public-data-sets-data-science-project/}}. Often, these data sets are community-created\footnote{\url{https://github.com/caesar0301/awesome-public-datasets}}. However, due to the fact that we aim at considering context events, i.e., events not directly related to the core business process, the set of usable data sets is significantly narrowed. 

In the following, we briefly show the most promising data set candidates and discuss our selection.
We have identified a number of possible data sets such as the data from Bosch Production Line Performance\footnote{\url{https://www.kaggle.com/c/bosch-production-line-performance}}. In the underlying scenario of this data set, parts move through the production lines of a manufacturer. While this happens, features are measured and recorded. The data set has been anonymized with respect to the names of the features. For each measurement, the part is evaluated, and if it fails quality control, this is recorded and the part is dismissed. The data set is highly imbalanced with regard to the actual class (i.e., failures or passes), and contains a very high number of features ($>12,000$). While this data set has the advantage of being rather large, the drawback is that no related business processes are defined. Furthermore, due to the per-part nature of the data, no business process can be mined as neither a stream of events, nor temporal correlation between measurements are available. 

Another prediction-centered data set stems from the Transport and Logistics Case Study (Cargo 2000)\footnote{\url{http://s-cube-network.eu/c2k/}} . The latter includes events related to messages sent within Cargo 2000 \cite{cesana2000cargo} (now known as Cargo iQ\footnote{\url{http://www.iata.org/whatwedo/workgroups/Pages/cargo2000.aspx}}). As shipments travel through segments of their transport (e.g., transfers between flights, airlines), their routes are recorded. This data set has been sanitized with respect to erroneous or incomplete messages. The business process related to this data set has been already used in research~\cite{Feldman:2013} and might have been relevant to our approach. However, despite the inter-organizational context, the provided data set as well as the corresponding process are solely related to the freight forwarding company and not to the entire process collaboration. Neither the private processes of the other involved organizations nor their respective data are available.  
Further, as the data set has been sanitized, it is difficult to identify failures to predict.

A third candidate for our evaluation was the Commodity Flow Survey (CFS) data set\footnote{\url{https://www.census.gov/econ/cfs/pums.html}}. It contains data about 4.5 million shipments. The data describes various attributes of the shipments, e.g., the source and destination of the shipment, type of commodity, whether or not the shipment requires temperature control during transportation, value, weight, modes of transportation, or hazardous materials. However, similar to the Bosch data set, it consists of a series of single, independent entities. From this, it is difficult to create an actual event stream, since no temporal relation is given between the data entries.

Finally, the fourth candidate is taken from the BPI Challenge data sets. From this collection, various data sets have already been used in research, e.g., \cite{conforti2017filtering,evermann2017predicting}. We regard the BPI Challenge 2017 data set\footnote{\url{https://data.4tu.nl/repository/uuid:5f3067df-f10b-45da-b98b-86ae4c7a310b}}. It consists of an event log stemming from a Netherlands-based financial institute issuing personal loans to applicants. The process from which the data stems is not explicitly defined in the data set, but can be mined using a process miner; e.g., ProM~\cite{Rozinat:2006}.

\begin{table}[!t]
\begin{center}
\caption{Evaluation of Data Sets, Fulfillment (+), No Fulfillment (-), and Partial Fulfillment (+/-) of Criteria: (i) Publicly Available, (ii) Event-based, (iii) Correlated with Business Process, (iv) Documented, (v) Containing Context Events and (vi) Failures, and (vii) Inter-Organizational.}
\label{tab:datsetselection}
\begin{tabular}{|c|c|c|c|c|c|c|c|}
\hline
& (i) & (ii) & (iii) & (iv) & (v) &  (vi) & (vii)  \\
\hline
{Bosch} & + & + & - & + & - & + & - \\
\hline
{Cargo 2000} & + & + & + & + & - & - & +/- \\
\hline
{CFS} &+ & + & - & + & - & - & - \\
\hline
{BPIC 2017} & + & + & + & + & - & - & - \\
\hline
\end{tabular}
\end{center}
\label{default}
\vspace{-0.3cm}
\vspace{-0.3cm}
\end{table}%

\begin{figure*}[ht]
	\centering
	\includegraphics[width=0.78\textwidth]{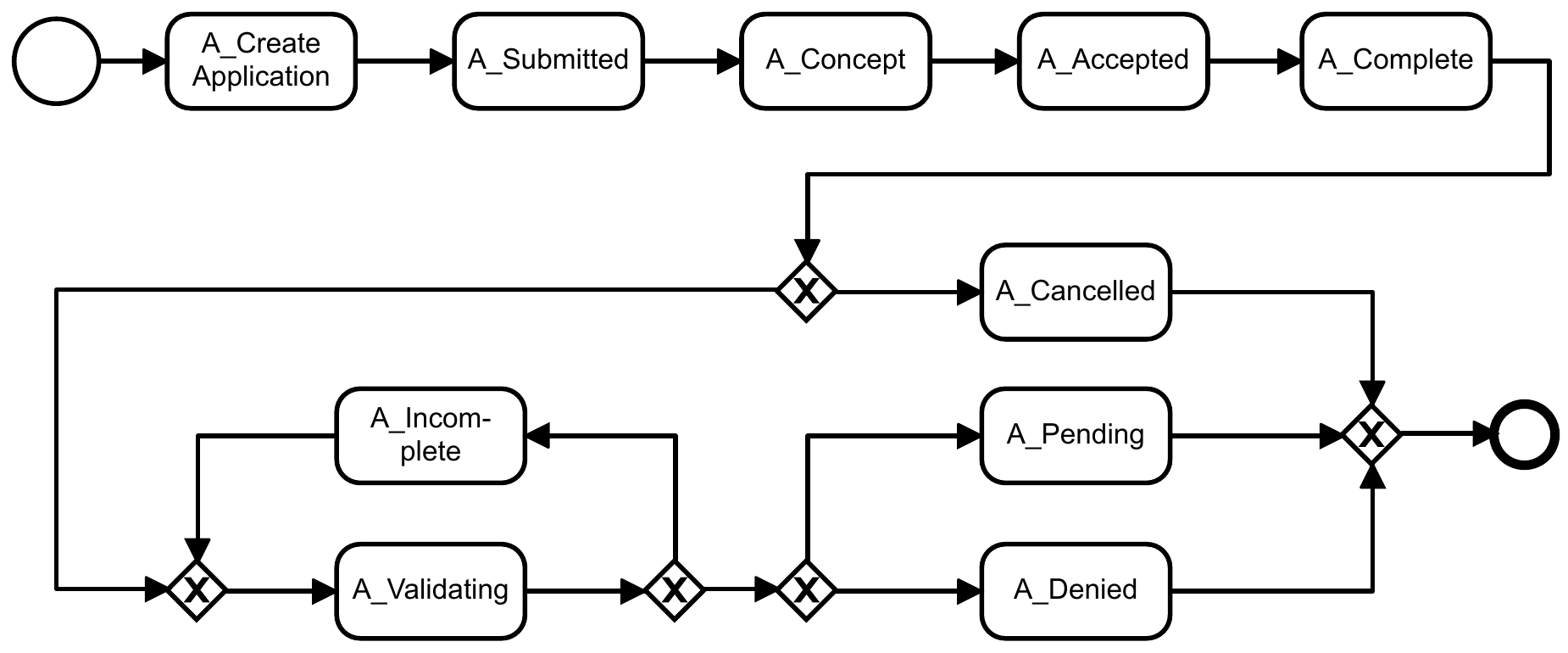}
	\caption{Process Model Mined from Real-World Data Set.}
	\label{fig:processmodel}
\end{figure*}

Table \ref{tab:datsetselection} summarizes the comparison of the data sets with respect to the specified criteria. Overall, no data set satisfies all criteria, but the Cargo 2000 and the BPI Challenge 2017 represent good candidates. The Cargo 2000 data set has (+/-) for criteria (vii) because despite the claim of supporting distributed processes, only data of one single organization process is provided. This puts it on the same level as the BPI Challenge data set. Overall, we have selected the BPI Challenge 2017 data set, as it satisfies most of the aforementioned requirements, with exceptions related to (iv) no explicit context events and (v) failures are provided in the data set, and (vii) it is not based on a distributed business process. 
We tackle the two first limitations by enriching the data set, albeit without compromising its relevance for real-world scenarios. A detailed description is given in Section~\ref{sec:prep-real}. For the third limitation, we decided to address the issue by using a synthetic data set generated from an artificial, but realistic process collaboration. It was inspired by the CFS and the Cargo 2000 data sets. Details about its generation are given in Section~\ref{sec:prep-synth}.
To conclude, we have opted for two data sets: (i) one from a real-world data source (BPI Challenge 2017) to prove the applicability of the approach, and (ii) an artificial data set to show its feasibility in a distributed setting.

\subsubsection{Real-World Data Set Pre-Processing}
\label{sec:prep-real}
Due to the limitations of the real-world BPI Challenge 2017 data set with regards to criteria (v) and (vi), data pre-processing is required. The main pre-processing involves the fact that the data set is an event log of a business process, and thus contains solely process-intrinsic events (i.e., no context events). Also, for our approach, the corresponding business process models are required. However, as no process model is provided along with the event log, applying data mining techniques is necessary. As described before, the data log is taken from the application process for personal loans from a Netherlands-based financial institution. The log consists of three different types of events: \emph{application} state changes, which have event names starting with \texttt{A\_}, \emph{offer} state changes, starting with \texttt{O\_}, and \emph{workflow} events, starting with \texttt{W\_}. An application within the process may contain one or more offers. One of the offers occurring within an application may be accepted. In this case, the entire application process is finished. However, if no offer is accepted, the application process is canceled. The application and offer events represent this process. The process events represent the necessary actions to be taken within the financial institution.

Since the core business process is the application for loans, we selected all application events (\texttt{A\_}) as process events. All remaining events (\texttt{O\_} and \texttt{W\_}) were used as context events. From the application events, we created a business process by using the Inductive Miner technique \cite{Leemans2013}. Using this technique is common in the field of process mining, and guarantees a certain level of \emph{rediscoverability} \cite{hernandez2015handling}. We used the ProM tool\footnote{\url{http://promtools.org/}} for process mining. The resulting process model is shown in Figure~\ref{fig:processmodel}. Furthermore, failure injection was also necessary to measure the accuracy of the prediction. Details about failure injection will be presented in Section \ref{sec:faults}.

\subsubsection{Synthetic Data Set Pre-Processing}
\label{sec:prep-synth}
As described in detail in Section~\ref{sec:sources}, in addition to our first, real-world data set, we use a second, synthetic data set. We opted for the generation of an artificial but realistic process collaboration. Data generation was inspired by the aforementioned CFS data set. This data set contains information on domestic freight shipments in different domains such as manufacturing and wholesale. Data include type, origin, destination, transport mode and other shipment attributes. 
The data set, however, includes one single event type rather than a stream of different event types, with neither time stamps nor correlation to process tasks or partners (no cases nor traces). 
Therefore, we have defined a collaborative process example of a supply chain scenario where goods are ordered, manufactured and shipped to the end client \cite{7207349}.  Process models details and description can be found in supplementary material\footnote{\label{supp}\url{http://gruppe.wst.univie.ac.at/projects/crisp/index.php?t=ebsdbpm1}}. The scenario involves six process partners, i.e., a bulk buyer, a manufacturer, two suppliers, a special carrier and a middleman. 

For each process partner, \emph{private} and \emph{public} tasks as well as interactions (through message exchanges) were defined. For each interaction, an XML template specifying the data elements to be exchanged has been created. The latter ensures \emph{consistency} of data instances for the simulation. Indeed, within one execution of the entire process collaboration, the data required by one partner task might depend on the output or data of another partner task. Therefore, it is important that the generated data is consistent, even though it is produced randomly. For instance, the delivery date of an item by the \emph{carrier} must not exceed the delivery deadline specified by the \emph{bulk buyer} and transmitted to the \emph{manufacturer}. In total, the collaboration contains 48 tasks distributed over the partners, of which 15 are interactions. Also, 20 data instances of message templates have been generated. 

The process collaboration was simulated using the Cloud Process Execution Engine (CPEE) \cite{5410270} in a \emph{distributed} way, where each process partner was executed separately on a different CPEE instance. The CPEE was chosen because it provides an efficient, flexible and lightweight way of executing distributed workflows, while its modularity allows us to collect the events.
An asynchronous correlation mechanism was implemented, which correlates the messages of different partners. To this purpose, a global instance identifier has been defined and exchanged through messages. The latter is primordial for process partners to correlate a received message with the correspondent process instance. We also distinguish between a process instance and a \emph{collaboration instance}. While the former represents the instance of one single process, the latter refers to one execution of the entire collaborative process.  

\begin{figure*} [ht]
	\centering
	\includegraphics[width=0.8\textwidth]{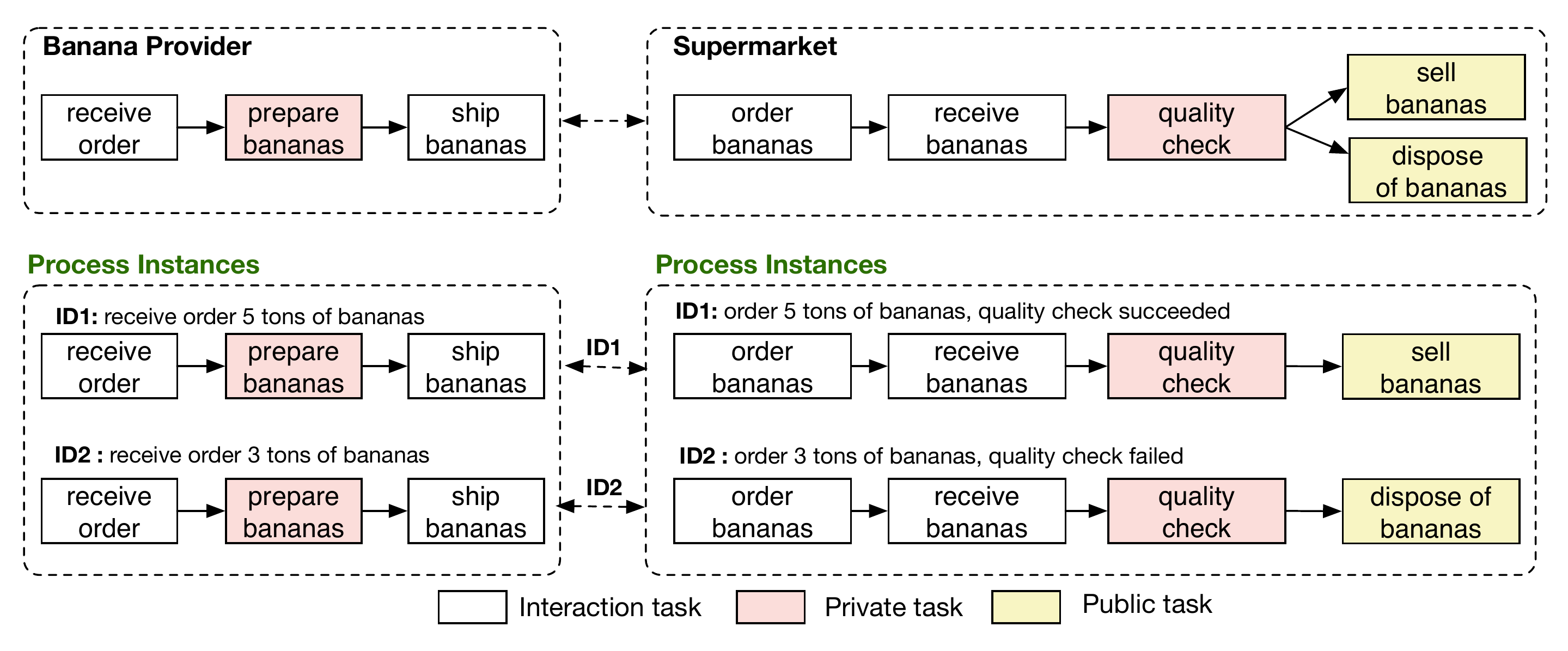}
	\caption{Collaborative Process Instances Used in the Synthetic Data Set.}
	\label{fig:instances}
\end{figure*}

For example, Figure~\ref{fig:instances} describes an example of two interacting processes, a banana provider and a supermarket. Each process is executed multiple times and each instance of the banana provider process must be correlated with the corresponding instance of the supermarket process. The instance identifiers $ID1$ and $ID2$ are specified in all message exchanges to ensure that data of one partner instance will be consumed by the right partner instance. A correlator (not shown in the figure) associates a message with the corresponding process and task instance (e.g., a message \emph {5 tons banana order} with task \emph{receive order} of instance id $ID1$). The latter can be centralized or distributed.    

The CPEE enables an easy collection of events including control and data flow as well as transactional information about the tasks, e.g., information about a task finishing or aborting. All events are stored in an Extensible Event Stream~(XES) file~\cite{xes}. The mechanism used to collect execution data from the CPEE also allows an easy collection and integration of events sent by a context event provider other than the CPEE~\cite{cs4790}. This can be important for the prediction, as some process tasks might depend on a context event provider, e.g., sensor sources external to the CPEE. With respect to the previous example, Listing~\ref{lst:xes} shows an example of an event of type \emph{order\_banana} from the automatically generated XES file. The full XES file can be found in supplementary material\footref{supp}.

\begin{lstfloat}[t]
\begin{lstlisting}
<log xmlns="http://www.xes-standard.org/" xes.version="2.0" xes.features="nested-attributes">
<extension name="Time" prefix="time" uri="http://www.xes-standard.org/time.xesext"/>
<extension name="Concept" prefix="concept"uri="http://www.xes-standard.org/concept.xesext"/>
<extension name="Organizational" prefix="org"uri="http://www.xes-standard.org/org.xesext"/>
<extension name="Lifecycle" prefix="lifecycle" uri="http://www.xes-standard.org/lifecycle.xesext"/>

<trace xmlns="http://www.xes-standard.org/">
 <string key="concept:name" value="Instance 487"/>
 <event>
  <string key="concept:name" value="order banana"/>
  <string key="concept:instance" value="http://cpee.org/~demo/corr/corr.php"/>
  <string key="id:id" value="a4"/>
  <string key="lifecycle:transition" value="complete"/>
  <list key="data_received">
   <string key="result">
    <order_banana>
     <id>m11_1</id>
     <quantity>2000 tons</quantity>
     <delivery_deadline>2016-12-30<delivery_deadline>
     <date>2016-12-18<date>
     <address>California<address>
    </order_banana> 
   </string>
  </list>
  <date key="time:timestamp" value="2016-12-15T15:58:37+01:00"/>
</event>
</trace>
</log>
\end{lstlisting}
	\caption{Example of an XES Event Log.}
	\label{lst:xes}
\end{lstfloat}

As a first step, we consider that all events are visible to all partners and therefore it is possible to analyze the entire XES file. Then we consider a more restricted view, where a partner can only see the data which it is allowed to see (its process, the public tasks of other partners and the interactions). The second scenario allows to consider the privacy aspects within a collaboration.

\subsubsection{Fault Injection}
\label{sec:faults}
As both selected data sets do not provide readily-available, explicit failures, modifications to the data sets are necessary. For the synthetic data set, we perform failure injection as described in this section. We have identified the following three major fault types to be injected into the synthetic data set.

\begin{description}
	\item[Step-indicated faults] For this type of faults, the process shows a sequence of steps which is characteristic for the given fault. For instance, a fault may cause an XOR gateway to proceed with a different process step than it would, had the fault not occurred.
	\item[Event-indicated faults] Faults manifesting themselves only through certain events, but not through different process steps executed. For instance, a sensor measuring the temperature of a container of bananas may sense the violation of certain temperature limits and fire an event. In our model, this corresponds to a context event being fired during the process execution.
	\item[Data-indicated faults] Certain faults are only indicated by the actual data associated with certain events. Considering the previous example, a temperature sensor may be recording temperature as events, regardless of whether a limit has been exceeded or not. In this case, the firing of the event itself does not correspond to a fault on its own. In fact, its associated data (the container temperature) is the determining factor of whether a fault has occurred.
\end{description}

It is noteworthy that these three types of faults exhibit an increasing level of difficulty for ML systems. While step-indicated faults can be detected by observing the executed steps (intrinsic events), event-indicated faults are harder to detect, because context events must be captured (i.e., selected as attributes for an ML model) and associated with the process model under execution. Finally, data-indicated faults require inspection of the data associated with events to detect an error and predict a failure. Naturally, for the detection of step-indicated faults, context events are not necessary, because only intrinsic events (representing business process step transitions) are taken into account. Nevertheless, since step-indicated faults are possible in reality, we also inject this type of faults, in order to gain insight into the system's reaction.

\begin{figure}
	\includegraphics[width=\columnwidth]{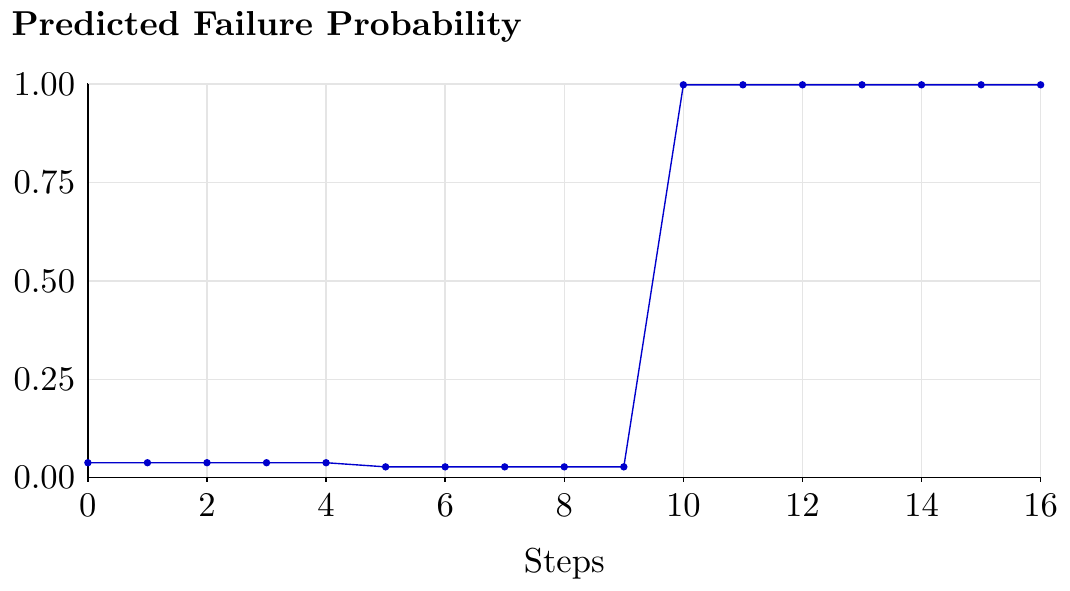}
	\caption{An Example of an Execution Timeline: The Error Occurred at Step~12, and the Predictor Immediately Determined an Imminent Failure for Step~16.}
	\label{fig:timeline}
\end{figure}

We inject faults randomly, using a given \emph{fault injection rate}. This rate is varying, and part of our parameter sensitivity analysis in the later parts of this paper. We randomly select one of the three aforementioned fault types when injecting a fault. According to fixed fault-error-failure combinations, we add or change the corresponding events to reflect the errors, and measure the system's ability to properly predict failures stemming from these errors. Following this, we feed the event streams with injected faults through the ML predictor component, and, for each injected fault, record whether and at which point in time the failure is predicted. Figure \ref{fig:timeline} shows an example of such an evaluation run, showing that the injection of a fault which is exposed as an error at step 12 is detected by the predictor. The predictor then changes from indicating almost certain success (0.0276) to almost certain failure (0.9980).

The real-world data set did not require such an amount of pre-processing. In the BPI Challenge 2017 data set, around 35\% of all loan application processes ended with the process step \texttt{A\_Cancelled} (as mined from the event log). This means that the application was cancelled, e.g., because of missing information or withdrawal by the applicant. Since this represents a failure to finalize the loan, we defined this state as a failure, in other words, all loan processes ending in this state were treated as failed. No further injection was required for the real-world data set.

\subsection{Experiments}
\label{sec:experiments}
The goal of our evaluation is to show that deploying our EFP component to the two selected data sets indeed yields useful prediction results. As described in Section~\ref{sec:related}, the approach presented in the paper at hand is novel in integrating external data sources and a common data format, while using an approach based on ML and taking into account the visibility of event data in an inter-organizational settings. Since a reference baseline to compare our results to is not available, we perform extensive experimentation to show the feasibility and applicability of our approach, and to motivate and provide a baseline for future research.

Our experiments all involve running an instance of our EFP component, feeding the evaluation data sets into it, and recording its performance in predicting the failure for probabilities. For initial experimentation and tuning, we use a fixed split of $70:30$ between training and test sets. The final results shown in this work, however, were obtained using 10-fold cross validation. The results of the 10 runs of the cross validation are averaged, with their standard deviation provided together with the arithmetic mean~\cite{kohavi95}. This is a widely accepted standard procedure in the evaluation of prediction solutions.

We derive confusion matrix metrics, i.e., True Positive~(TP), True Negative~(TN), False Positive~(FP), and False Negative~(FN)~\cite{manning09}. In our evaluation, \emph{positive} denotes the presence of failure, and \emph{negative} denotes the absence of a failure in an event trace. From these values, we derive three metrics: precision, recall, and the Matthews correlation coefficient (MCC). All three are common metrics for evaluating binary classification algorithms \cite{Sokolova2006,mcc1,mcc2}.

The definitions of precision and recall are shown in (\ref{eq:p}) and (\ref{eq:r}), respectively. Precision determines the fraction of correctly-classified positive (failure-containing) instances, relative to all positively classified instances (``Out of all \emph{fail} classifications, how many instances actually contain a failure?''). Recall determines the fraction of correctly-classified positive instances, relative to all \emph{actually} positive instances (``Out of all actual failures, how many instances are marked as \emph{fail}?'').

The MCC is an application of the Pearson correlation coefficient. We use a variant of the MCC deriving the values directly from the confusion matrix, as shown in (\ref{eq:mcc}) -- this representation is equivalent to the original defined in \cite{mcc1}. An MCC of 1 denotes total positive correlation, an MCC of 0 indicates no correlation, and an MCC of $-1$ denotes total negative correlation. The value of the MCC determines the measure of linear dependence between the actual outcome (failure or success) and the prediction. While precision and recall give a good classification performance overview, the MCC provides balance between the positive and negative class instances.
\begin{align}
	\text{Precision} &= \frac{ \text{TP} }{ \text{TP} + \text{FP}} \label{eq:p}
\end{align}
\begin{align}
	\text{Recall} &= \frac{ \text{TP} }{ \text{TP} + \text{FN}} \label{eq:r}
\end{align}
\begin{align}
	\text{MCC} &= \frac{ \text{TP} \times \text{TN} - \text{FP} \times \text{FN} }{ \sqrt{ (\text{TP} + \text{FP})(\text{TP} + \text{FN})(\text{TN} + \text{FP})(\text{TN} + \text{FN}) } } \label{eq:mcc}
\end{align}

Our first experiment aims at verifying the overall performance of our approach. For this, we use the real-world data set described in Section~\ref{sec:prep-real} as input for our EFP component. The resulting confusion matrix is shown in Table \ref{fig:confusion}, and the mean values of the metrics are summarized in Table~\ref{fig:metrics}.

\begin{table}
	\centering
	\footnotesize
	\caption{Confusion Matrix for Evaluation Using Real-World Data Set (Mean Values for 10-fold Cross Validation, all $\sigma < 0.4$)}
	\begin{tabular}{c|l|c|c|c}
		\multicolumn{2}{c}{}&\multicolumn{2}{c}{Classification}&\\
		\cline{3-4}
		\multicolumn{2}{c|}{}&Positive&Negative&\multicolumn{1}{c}{$\Sigma$}\\
		\cline{2-4}
		Actual & Positive & $1051.14$ & $28.04$ & $1079.18$\\
		\cline{2-4}
		Class & Negative & $153.76$ & $1917.95$ & $2071.71$\\
		\cline{2-4}
		\multicolumn{1}{c}{} & \multicolumn{1}{c}{$\Sigma$} & \multicolumn{1}{c}{$1204.90$} & \multicolumn{    1}{c}{$1945.99$} & \multicolumn{1}{c}{$3150.89$}\\
	\end{tabular}
	\label{fig:confusion}
\end{table}
\begin{table}
	\centering
	\footnotesize
	\caption{Key Metrics for Evaluation Using Real-World Data Set (Mean Values for 10-fold Cross Validation)}
	\begin{tabular}{c|c|c}
		Metric & Mean & $\sigma$ \\\hline
		Precision & $0.873$ & $0.344$ \\
		Recall & $0.971$ & $0.307$ \\
		MCC & $0.879$ & $0.331$
	\end{tabular}
	\label{fig:metrics}
\end{table}

The results show a precision of $0.873$, a recall of $0.971$ and an MCC of $0.879$. It is also noteworthy that the standard deviation ($\sigma$) is relatively low. This, together with the cross-validation performed, shows that the classifier resulting from the ML model training performs steadily across the entire real-world data set, and that the performance is not dependent on which one of the partitions was used for training, and which one was used for testing.

In the second part of our experimentation, we seek to determine the impact of the amount of data available for prediction. In other words, we are interested in how much the visibility of private events impacts the performance. This requires using the synthetic data set. Since within this data set, we inject failures with a given rate, we are interested in the impact of this injection rate parameter on the results. We therefore analyze the sensitivity to this parameter in order to avoid bias stemming from parameter choice.

\begin{figure}[t]
	\includegraphics[width=\columnwidth]{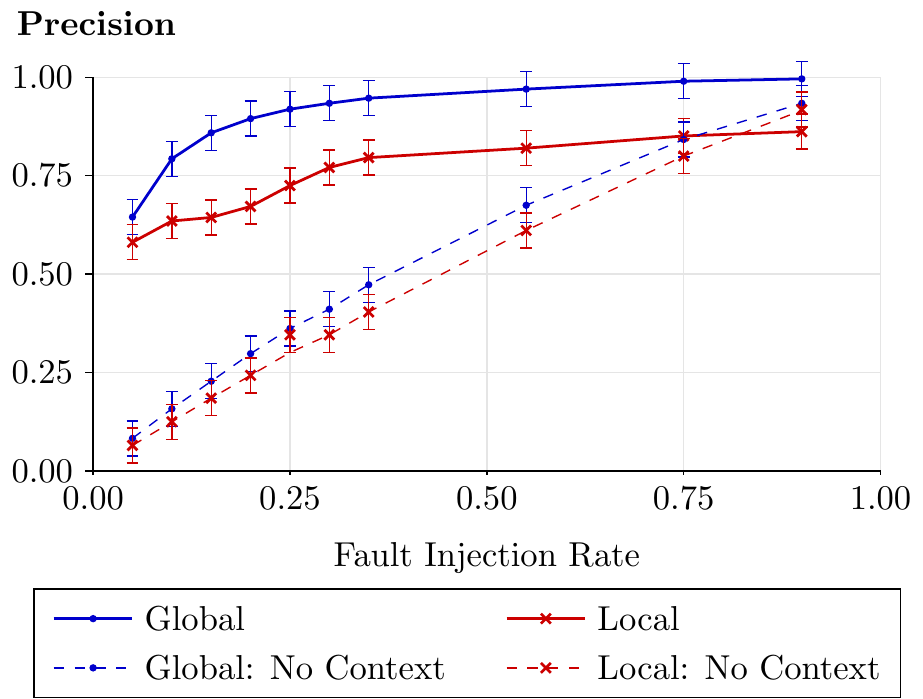}
	\caption{Classifier Precision Using Synthetic Data Set, over Varying Fault Rates.}
	\label{fig:synth-pre}
\end{figure}

\begin{figure}[t]
	\includegraphics[width=\columnwidth]{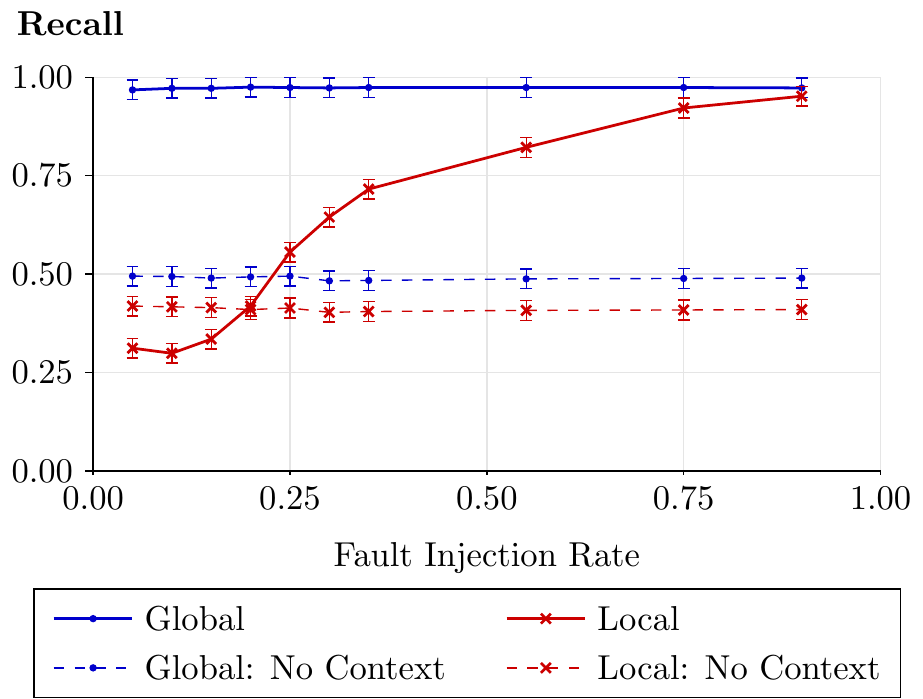}
	\caption{Classifier Recall Using Synthetic Data Set, over Varying Fault Rates.}
	\label{fig:synth-rec}
\end{figure}

\begin{figure}[t]
	\includegraphics[width=\columnwidth]{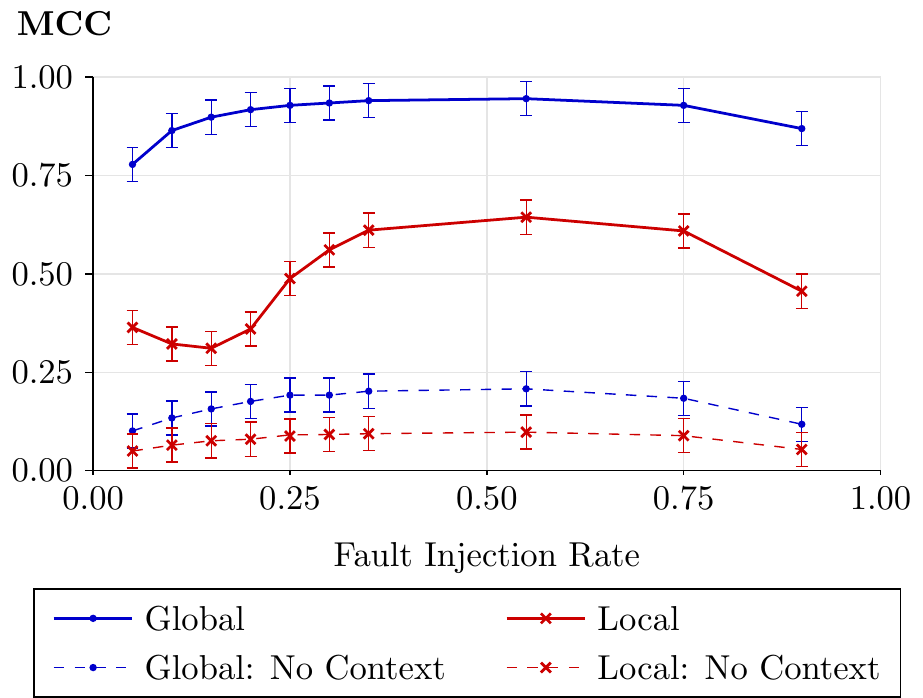}
	\caption{Classifier MCC Using Synthetic Data Set, over Varying Fault Rates.}
	\label{fig:synth-mcc}
\end{figure}

We show the results for precision, recall and MCC in four different scenarios. In the \emph{global} scenario, events from all process partners are available to the EFP component (this corresponds to having all events marked as public events). In contrast, the \emph{local} scenario only provides the EFP component with events from a single partner. This represents the privacy scenario from the use case described in Section~\ref{sec:prep-synth}. From these two scenarios, we derive two further scenarios, leaving out the context events, providing a baseline to compare the results to. All values are provided together with their standard deviations $\sigma$, marked using error bars. The results for precision are shown in Figure~\ref{fig:synth-pre}, recall is shown in Figure~\ref{fig:synth-rec}, and the results for MCC are shown in Figure~\ref{fig:synth-mcc}. The intuition that less available data decreases classification performance is clearly visible. Nevertheless, the data gives insights into the amount of performance decrease caused by only using the data from a specific partner in the evaluation, and also provides a comparison of results between scenarios with context (\emph{global} and \emph{local}) and without context events (\emph{no context}). The biggest drop in performance is seen for recall at a fault rate of $0.10$, where the \emph{local} scenario reduces the recall value from $0.972$ to $0.299$. Removing context events generally decreases precision and recall and therefore also the MCC, except for the corner cases of very low fault rates (where both \emph{no context} scenarios have better recall values than \emph{local}), and very high fault rates (where both \emph{no context} scenarios have better precision values than \emph{local}). In any case, the \emph{global} scenario shows significantly better results than \emph{local} and \emph{no context} across the entire evaluation domain.

Furthermore, analyzing the impact of varying fault injection rates in the synthetic data set, we see that while precision and recall generally increase with a higher amount of faults, the MCC shows a \emph{sweet spot} around $0.50$, for all three executed scenarios. This knowledge, however, has no universal application, since the fault rate is highly domain-specific. Comparing this data to the real-world data set, however, shows that the results at the fault rate of the real-world data set ($35\%$) yields results comparable to the \emph{global} scenario (since the real-world data set only allows this scenario).

\subsubsection{Discussion, Implications and Limitations}

Generally, we observe an increase in both precision and recall for an increasing fault rate. While it is difficult to determine a reason for certain performance metrics when ANNs are involved, we suspect that the types of processes used in the data set cause a high failure rate to be easier for ANNs to process than low failure rates. The \emph{global} scenario generally shows better recall than precision, i.e., a low number of FNs. In contrast, the \emph{local} scenario shows higher precision values than recall, especially for low fault rates. We suspect that the inter-organizational structure of the process involved in the synthetic data set benefits an accurate detection of inherent faults, increasing the precision. This, in turn, comes at the cost of a lower recall.

The \emph{no context} scenarios aim at giving a baseline for our approach by not taking into account context events at all.
We can observe this scenario to be superior to \emph{local} for high fault rates ($> 0.75$) with regard to precision, and for low fault rates ($< 0.20$) with regard to recall. These figures represent some noise that context events can add that distort the prediction for those extreme cases of fault rates. The MCC, which aims to find a one-metric balance between many possible metrics to compare classifiers, however, is consistently higher for \emph{local} than for \emph{no context}, and the highest for \emph{global}.
From these results, we conclude that considering context events clearly yields drastic improvements in precision, recall and MCC values.

It is worth pointing out that the distinction between intrinsic and context events is made on a conceptual level in this approach, but the evaluated data sets do not fully reflect this distinction. This stems from the fact that, as described in detail in Section~\ref{sec:sources}, finding a data set stemming from a real-world source and containing proper context events proved to be difficult, and at the same time, while the synthetic data set contains both intrinsic and context events, both are generated, so the distinction might indeed seem blurred. Furthermore, the failures in the real-world data set were generated by treating a certain event (\texttt{A\_Cancelled}) as a failure. This selection poses a limitation to our evaluation, since the processes described in the data set might have also failed in other ways, not generating those events. Nevertheless, we regard those measures as necessary for the evaluation of our approach, and the experiments confirm that using context events as an additional data source in predicting failures is viable.

Finally, we note that while our approach does not require experts to create rule-based failure prediction models, since ML is used, it still requires certain expert input: (i) The creation of a suitable ML model still requires human intervention for selecting ML models and tuning parameters, (ii) the inclusion of certain data sources as external events also requires expert knowledge. However, we argue that the area of expertise required is different. While for creating a rule-based prediction model, experts in the given domain are necessary, in our case, the required knowledge is more abstract, as skills in ML are required, and domain-specific knowledge is less relevant.

\subsubsection{Evaluation Result Summary}

The proposed approach yields satisfactory results for predicting failures in business processes. The resulting metrics show a precision of $0.873$ ($\sigma = 0.014$), a recall of $0.971$ ($\sigma = 0.008$) and an MCC of $0.879$ ($\sigma = 0.012$).
In the second part of our evaluation, we analyze the performance when facing a scenario where multiple process partners collaborate, but do not share all of their events, which is possible in situations where privacy aspects prevent a business process partner from sharing internal events.
We observe that the impact on the performance is measurable, and, depending on the metric and fault rate, can decrease the prediction performance from around $0.972$ to $0.299$ at a fault rate of $0.10$.

It is noteworthy that this is a parameterizable approach, i.e., the exact learning model differs from domain to domain. Therefore, while our approach reduces the necessary expert knowledge required to create failure prediction models, there is still the need for an expert to tune the learning parameters to achieve satisfactory results. Nevertheless, our approach can provide a flexible and generic methodology of predicting failures in business process execution.

\section{Related Work}
\label{sec:related}

The usage of events in BPM has gained some attention in recent years, with a focus on fields like complex event processing (CEP) for business processes, business process intelligence~(BPI), and business activity monitoring (BAM)~\cite{janiesch12, krumeich14}. 
Some work has been done in the area of failure prediction and fault tolerance, as will be discussed in more detail below\footnote{A more extensive discussion of event-based BPM can be found in~\cite{krumeich14, schulte15}.}.
Nonetheless, only few approaches consider the execution context of BPM, thus exploiting the presence of context events in combination with those 
generated by the BPM execution.

\textbf{Integration of BPMS and EBS.}
The integration of BPMS and EBS has led to the development of new integrated solution aimed to perform several key tasks: distribution, control, monitor, and predictive analysis for business processes.

Event-based BPM relies on the principle of event-driven architectures (EDAs), which describe an architectural style with event-driven components and communication~\cite{luckham02}.
EDAs resemble technical aspects of publish-subscribe middlewares~\cite{eugster03}, especially regarding the decoupling of components, and the pushing of events~\cite{buchmann12}.
In event-based BPM, an EDA allows to communicate events between different process stakeholders, components, and process steps, and therefore to control and change business processes during runtime and design time~\cite{vonAmmon:2009}. 
In the process of distributing the execution of BPMS, e.g.~\cite{Li:2010},  %
several works, e.g.,~\cite{jacobsen:2015},~\cite{jacobsen:2016}, exploit the decoupling properties of publish-subscribe systems, 
to allow the distributed execution of business processes. 
The very idea is to exploit the loosely coupled and distributed nature of publish/subscribe systems. 
The BPMS becomes an event sink as well as an event source, thus generating and consuming process-related events~\cite{etzion10,schulte15}. 

Importantly, the BPMS can be controlled by an EBS through events~\cite{janiesch12}.
In an early approach to use events in BPM, von Ammon~et~al.~\cite{vonAmmon:2008} present a basic reference model to control processes. Event types from BPMN 1.1 are supported, which includes exception events. 
The main focus of this work is on a generic approach to use events to control a process instance, hence the authors do not discuss how exception events could be generated or how to derive failures from events. 
An example for the usage of CEP in order to adapt a business process instance is presented by Hermosillo et al.~\cite{hermosillo10}. The authors propose to adapt a process at runtime based on predefined, event-based rules, which makes the approach rather inflexible. The main contribution of this work is a language to define these rules and when and how to apply them; nevertheless, there is no discussion on the nature of the events and how to identify a critical event if it has not been specified in a rule. A related approach has been presented for scientific workflows~\cite{zhao12}. Here, event messages are emitted by distributed agents for workflow execution control. Only events explicitly generated by agents are taken into account, i.e., context events are not explicitly regarded. Reactions to events are done based on predefined rules.

Several authors have proposed event-driven monitoring approaches, e.g.,~\cite{DiFrancescomarino:2017, Feldman:2013, schwegmann13}. 
For instance, Feldman et al.~\cite{Feldman:2013} show an effective example of event-based prediction. In their use case, real-time monitoring data is used to anticipate issues of cargo shipments. 
The prediction model relies on a simple stochastic approach, thus it is able of discovering only direct relations between system state and predicted outcome. 
More sophisticated approaches for prediction can be investigated, like Schwegmann et al.~\cite{schwegmann13} do.
The authors combine real-time event monitoring with predictions of future process behavior. The resulting model allows the usage of a number of ML-based predictors, similarly to our approach. 

Our work combines a BPMS with an EBS aiming to monitor the execution of BPM and perform predictive analysis. 
We are not interested in discovering complex events, but rather in processing the huge amount of BPM-related data to predict the future system evolution.
Indeed, similarly to the work by Schwegmann et al.~\cite{schwegmann13}, we rely on an ML approach to predict failures. 
However, differently from all previously discussed work, we augment the events generated by the BPMS running the business processes with context-related events; 
the latter can encompass events generated by IoT devices (e.g., temperature or position sensors) placed in the real-world execution environment of the business process. 
By distinguishing between intrinsic and context events, we strengthen separation of concerns. This, in turn, allows for a better reuse of the shared context among multiple business processes.

\textbf{Failures and Their Prediction.} 
Apart from the approaches discussed so far, which focus on generic process adaptation and BAM, there is also a number of approaches explicitly aiming at fault tolerance for business process executions. From a technical point of view, different strategies how to make service compositions fault-tolerant have been proposed~\cite{zheng15}. 
Fan et al.~\cite{fan12} introduce a fault tolerance strategy for service compositions which includes failure detection. However, the detection is done by comparing an expected result with the actual outcome of a service composition. Hence, the approach is rather inflexible and requires modeling of the expected results; an actual failure prediction is not carried out.
Events are generally not taken into account, only the outcome of a service composition is assessed. 
In addition, a large number of approaches to tolerate \emph{non-functional} faults (e.g., delays) in service compositions have been proposed, e.g., with a focus on recovery mechanisms~\cite{console08}. For instance, Leitner et al.~\cite{leitner13a} use ML techniques to predict performance faults. In an earlier approach, Canfora et al.~\cite{canfora05} propose to re-plan service compositions during runtime based on the actual QoS. The approaches discussed in this paragraph focus on the technical level of service compositions, not taking into account context events as observed in the work at hand, but rather aiming at failures arising from the execution of software-based process steps.

Failure prediction for business processes is also partially related to the field of anomaly detection in process executions. For instance, Bezerra et al.~\cite{bezerra09} use process mining in order to classify anomalous and normal instances of a particular process model. The outcome of this anomaly detection is an {\em ex post} analysis whether something uncommon did happen. Also, the approach does not implicitly take into account events. 
Instead, process logs are mined. Hence, only anomalous process steps are taken into account, while we argue that context events actually may precede such steps at runtime.

To grasp the complex relation among system components and to discover early symptoms of failures to come, several works rely on ML techniques, e.g.,~\cite{Yoo:2016,Teinemaa:2016,Leontjeva:2015,Kang:2012,AbuSamah:2017,Salfner:2007}.
Abu-Samah et al.~\cite{AbuSamah:2017} rely on Bayesian networks; as a drawback, this approach requires to be complemented with the extraction and validation of system patterns, which may involve expert opinions or elicitations on several levels.
Leontjeva et al.~\cite{Leontjeva:2015} address the problem of predicting the (positive or negative) outcome of an on-going business process by analyzing event logs using a Hidden Markov Model. As such, the solution assumes the Markovian property, therefore it cannot easily take into account long-term dependencies among events and outcome, like we do.
An approach based on Hidden Semi-Markov Models, which loosen the Markovian property, has been presented in~\cite{Salfner:2007}. 
Pika et al.~\cite{Pika:2013} provide a solution based on statistical analysis aimed to identify the risk of deadline overruns of processes.
A more flexible solution is proposed in~\cite{Kang:2012}, where Kang et al. aim at the detection of abnormal process \emph{termination}, and, similarly to our work, the authors apply ML to achieve real-time fault detection. 
However, the authors focus on process-intrinsic knowledge, while context events are not taken into account. In the end, their approach compares the actual process execution with the expected process execution. Nevertheless, this work comes closest to the work at hand. 
Although focused on process events only, an interesting approach is proposed 
by Teinemaa et al.~\cite{Teinemaa:2016}; it jointly exploits unstructured (free-text) and structured data to predict the process outcome. Even though we do not consider unstructured data, our approach could embed the principles presented in~\cite{Teinemaa:2016}.

Grambow et al.~\cite{grambow12} provide an approach to event-based exception handling for processes. Importantly, the authors focus on software engineering processes, not on business processes in general. Their approach to identify critical events is based on the event-condition-action pattern, which makes it necessary to define events and conditions to handle them. While the authors claim that their approach is able to take into account unanticipated conditions, it remains unclear how this is achieved. Events are related to process activities and artifacts, while context event sources are not regarded.
In a more specialized approach, Pika et al.~\cite{pika16} aim at process risk management through the analysis of event logs. Since the authors focus on risk management aspects, their work exceeds the work at hand in terms of prediction of a process outcome: While we focus on failure prediction, Pika et al. also predict pre-defined possible process outcomes, e.g., timeliness of single process steps. To identify risks, process models are annotated with guards, while in our approach, we do not require such prearrangements. 
We refer to~\cite{Salfner:2010} for an extensive survey on online failure prediction methods.
As shown in~\cite{Metzger:2015}, 
 different prediction techniques differ in terms of accuracy and ability to capture specific phenomena. Hence Metzger et al. have proposed and evaluated the idea of combining several techniques, so to achieve better performance. This is surely an interesting idea, which could become part of our future work. 

Other works move in the direction of identifying root causes of failures. 
Conforti et al.~\cite{Conforti:2015} propose another approach aiming at process risk management based on the analysis of event logs. The goal of the authors is to minimize risks by identifying potential risks (e.g., timeliness, reputation, cost) during the scheduling of work items. 
An ML approach is applied on process-related events, thus neglecting context event sources. 
Examples for root cause analysis based on event data and using ML have been proposed before~\cite{Rozinat:2006,Suriadi:2013}. %
While we do not focus on root cause analysis for process faults, this could be another interesting direction for future work.

\textbf{Context-awareness.} 
Context-awareness helps to improve decision-making processes by introducing new information that can better describe the execution conditions of a business process. 
Albeit the importance of context information, only few works considers them explicitly, e.g.,~\cite{hermosillo10, 4595562, Bohmer:2016}.
In~\cite{4595562}, the authors investigate the most commonly used kind of context information employed so far (but not in the field of BPM).
The authors conclude by assessing that performing adaptation in response to changing execution condition, captured by context information, may be very beneficial to software systems. 
A similar idea results from the work in~\cite{hermosillo10}, which relies on events to control the BPM process execution.
Nevertheless, the latter proposes a rule-based approach which may not be flexible enough with regard to previously unknown context information. Conversely, we leverage on the flexibility of ML approaches to work in presence of concept drifts~\cite{widmer1996learning}. 

In~\cite{Bohmer:2016}, B\"{o}hmer and Rinderle-Ma provide a runtime approach to anomaly detection which also takes into account the context of a process, including time- and resource-related events. The main goal of the authors is to identify malicious attacks on process executions. The authors base their approach on an explicit set of expected execution events and their likelihoods of occurrence. 
In contrast, our approach only regards the expected execution events in an implicit way, thus leading to higher flexibility. %
Also, the approach presented by B\"{o}hmer and Rinderle-Ma focuses on a predefined set of event sources, while we allow the integration of arbitrary sources, through the concept of context events. 

In summary, the approach presented in this paper is novel since it proposes an integrated solution which takes into account the following key points which have not yet been fully regarded in the existing literature. 
First, most related work focuses on event logs. While a number of existing approaches claim that arbitrary events or events from the IoT could be taken into account, this is not explicitly foreseen. 
Through integrating external data sources and a common data format (i.e., the XES format) into our solution, we are able to achieve this. 
Second, the current state-of-the-art in failure prediction mostly depends on pre-defined rules or conditions when a particular failure will arise. Through the application of a ML-based approach, our solution is more flexible, however still needs a training set where particular process instances are labeled as failing (or not). 
Third, we do not restrict our approach to a particular goal, such as monitoring, process adaptation, or risk management, but aim at generic failure prediction. 
Fourth, to the best of our knowledge, there is currently no discussion on how the visibility of event data in inter-organizational settings influences predictions of process outcomes. We conduct such an analysis as part of our evaluation in Section~\ref{sec:evaluation}.

\section{Conclusion}
\label{sec:conclusion}

In contrast to traditional BPM solutions, where the toolset for managing BPM systems is focused on centralized, intra-organizational processes, today's business processes are highly distributed, with inter-organizational stakeholders and decentralized architectures, calling for novel methodologies of detecting and responding to unforeseen events and failures. In this article, we discuss the need for such methodologies, and present an approach for employing event-based error detection and failure prediction for business processes. We evaluate our approach using two data sets. The first data set is a real-world business process data set from the finance domain. The second data set is a synthetic data set, stemming from a choreography model involving several partners collaborating in a common business process. We demonstrate that the implemented failure prediction component is capable of detecting upcoming failures. In the real-world data set experiments, the failure prediction component exhibits a precision of $0.873$, a recall of $0.971$ and an MCC of $0.879$, with relatively low standard deviations.

The results of the presented work provide a promising and motivating base for further research in the field of integrating BPMS and EBS. While the need for expert knowledge is not completely eliminated, the presented approach requires less domain-specific expert knowledge than the usage of a fixed set of rules, both in setup and in rule maintenance. The necessary expert knowledge is shifted from the domain itself to tasks within the domain of ML, e.g., feature selection.

Several future research directions can be identified, including the investigation of efficient online learning techniques and the transition to a real production system. Due to the complex and hidden relations among events, the automatic selection of data sources represents a critical task for accurate failure predictions. Furthermore, the aspect of scalability has only been covered briefly in our approach, and remains an important factor, since the cost of time and computing power necessary to perform the prediction may play a role in certain scenarios. Sensitivity analysis, for instance for the search space limiting parameters, would be interesting. Another promising direction of research is the further identification and integration of modern event sources such as IoT or smarter cities. Furthermore, a production system might bring up the need for self-adaptation to mitigate, for example, the negative impacts of concept drifts.

\appendix
\section*{Acknowledgment}
The original idea to this work is a result of the GI-Dagstuhl Seminar 16341 ``Integrating Process-Oriented and Event-Based Systems''. 

This work is partially supported by the Commission of the European Union within the CREMA H2020-RIA project (Grant agreement no.~637066) and by the Vienna Science and Technology Fund (WWTF) through project \mbox{ICT15-072}.

\section*{References}
{\renewcommand{\section}[2]{}
\bibliographystyle{elsarticle-num} 
\bibliography{ms}}
\newpage

\end{document}